\documentclass[american,aps,pre,groupedaddress,notitlepage,11pt,floatfix,superscriptaddress]{revtex4-2}
\usepackage[utf8]{inputenc}
\usepackage{subfig}
\RequirePackage{fix-cm}
\usepackage[table]{xcolor}
\usepackage{graphicx}
\usepackage{braket}
\usepackage{float}
\usepackage{ragged2e}
\usepackage{array}
\usepackage{dcolumn}
\usepackage[colorlinks=true, citecolor=blue]{hyperref}
\usepackage{hhline}
\usepackage{amssymb}
\usepackage{amsmath}
\usepackage[utf8]{inputenc}
\usepackage{tikz}
\def\beq{\begin{equation}}
\def\eeq{\end{equation}}
\begin{document}
\title{Energy-gap modulation and majorization in  three-level quantum Otto engine}
\author{Sachin Sonkar} 
\email[e-mail: ]{sachisonkar@gmail.com}
\affiliation{ Department of Physical Sciences, 
		Indian Institute of Science Education and Research Mohali,  
		Sector 81, S.A.S. Nagar, Manauli PO 140306, Punjab, India} 
\author{Ramandeep S. Johal} 
\email[e-mail: ]{rsjohal@iisermohali.ac.in}
\affiliation{ Department of Physical Sciences, 
		Indian Institute of Science Education and Research Mohali,  
		Sector 81, S.A.S. Nagar, Manauli PO 140306, Punjab, India}

\begin{abstract} 
A three-level quantum system having two energy gaps
presents a nontrivial working medium for a quantum heat engine.
Our focus lies in understanding the constraints on
the ability to modulate these gaps relative to the
changes in probability distributions at the two given heat reservoirs.
It is seen that an Otto engine in the quasistatic limit is feasible if
at least one energy gap shrinks during the first quantum adiabatic
stage. We analyze operating conditions
under different variations of the gaps,
revealing that a definite majorization relation between the hot
and cold distributions serves as a sufficient criterion for the
engine when both gaps are shrinking. Further, Otto efficiency
is enhanced in case majorization holds.
On the other hand, majorization becomes a necessary condition
when only one of the gaps is shrinking.
 In the special case where one gap remains
fixed, majorization is both necessary and sufficient
for engine operation.
For an $n$-level system, we note that a well defined change in energy
gaps aligns with the majorization relation, thus characterizing the operation of the engine.
\end{abstract}	

\maketitle
	

\section{Introduction}

A 3-level system (3LS) in thermal contact with two heat reservoirs
was the first prototype \cite{Threelevelmaser}
for a quantum heat engine, consistent with the Second law of
thermodynamics. Following this
discovery and subsequent developments \cite{Alicki1979, linblad1976,
Sheehan2002, Scully2003,
Allahverdyan2004, Mahlerbook1, Allahverdyan2008}, the field of quantum thermodynamics has literally
picked up steam and seen rapid advances in recent years \cite{quan2007quantum,Myers_2022,Vinjanampathy_2016,Millen_2016,quan2009}. Quantum thermal machines explore the resourcefulness of quantum features
such as coherence and entanglement \cite{altintas2014,borris2017,Jaegon2022,Camati2019,zhang2007four,Niedenzu2015,Kamimura2022,Scully2003} for energy conversion purposes. Such machines
are now being experimentally realized across various platforms  \cite{peterson2019experimental,assis2019,netterahein2022,abah2012,
hubner2014,PhysRevB.101.054513,PhysRevB.94.184503,PhysRevResearch.5.L022036,PhysRevLett.128.090602,QOCinasupercoducting,Statistsanyone,PhysRevLett.125.166802,PhysRevLett.122.110601,EITQHE_2017,flywheel_2019,Ro_nagel_2016,Maslennikov_2019,Bouton_2021,PhysRevLett.123.250606}.
The performance analysis of quantum heat engines, refrigerators and heat
pumps, has thus emerged as a major sub-discipline
in this domain, where quantum Otto cycle has been widely
studied using a variety of working media and reservoir
configurations---both in quasi-static
as well as finite-time regimes \cite{Altintas2015,Thomas2011,2014friction,venu2021,Sachin2023,Ivanchenko2015,turkpencce2019coupled,Brunner2014,zhang2019optimization,de_Assis_2021,Singh_2020,wwang2019,Huang_2017,wong2012,PhysRevB.99.024203,osgur_2023,El_Makouri_2023,hasegawa_2023,Piccitto_2022,Solfanelli_2023,PhysRevE.106.014114,PhysRevResearch.5.023066,sangik2022,Thomas2018,
wu2014,
das2020,
chand2021,
Lee2020,Geva1992,
saha2023temperature,
feldmann2000performance}.
In the paradigmatic model of a two-level Otto engine   \cite{alicki2014quantum,kieu2004,kieu2005quantum,Mahler2007,alicki2014quantum,Beretta_2012,wong2012,Ghosh_2018,Thomas2018,Papadatos_2023},
average heat flow from hot to cold reservoir is ensured
if the equilibrium probability distribution at the cold
reservoir majorizes the corresponding distribution at
the hot reservoir.
Specific working conditions for multi-level systems \cite{Altintas2015,Huang_2017,venu2021, Sachin2023,Uzdin_2014,El_Makouri_2023,Niedenzu2015,TDEROliveira2020} are usually
difficult to pin down, except when the energy spectrum
changes in a specific way.  Some many-body
systems may yield to analytic treatments, see e.g.
Refs. \cite{PhysRevB.99.024203, Simmons2023, Jaseem2023, Williamson2024}.

In particular, the technique of majorization
\cite{Marshallmajorizationbook,TSagawa,Bhatia1996MatrixA,Buscemi2017,Joe1990MajorizationAD,2020,vedral2015,2016,RUCH1980222}
has been found to be useful in certain coupled-spins models \cite{venu2021, Sachin2023}. The connection between performance of quantum heat engine and majorization was noted in Ref. \cite{venu2021}. Subsequently, the concept of majorization was employed to characterize the performance of the spin-based quantum Otto engine (QOE) in Ref. \cite{Sachin2023}. The majorization criterion has important implications for state transformation in various resource theories \cite{10.5555/2011326.2011331,PhysRevLett.83.436,PhysRevA.91.052120,PhysRevLett.83.3566,Horodecki2003,Uttam2021,lostaglio2022,majorizationcomplete,PhysRevResearch.2.033455,2015,2013,thermalcone_2022,Bosyk_2019}. Moreover, majorization serves as a sufficient criterion for ordering the maximum work extraction from a finite quantum system when it is in an equal energetic state \cite{Allahverdyan2004,Mir2020}.
The concept proves valuable in deriving bounds on the ergotropy gap,
with implications for  witnessing and quantifying entanglement \cite{Mir2019,Joshi2024,puliyil2022} and applications in
models of quantum networks \cite{vaishak2023} and quantum batteries \cite{MirAli2019}.

In this regard,
a 3LS provides a convenient platform to examine the influence of population and coherence on the performance of a thermal machine, avoiding any contributions from quantum entanglement or quantum correlations between subsystems.
A 3LS or a qutrit has important applications in quantum information processing
\cite{Fedorov2012, Yurtalan2020, Morvan2021}.
It is proving as a testbed for the study of heat transport and other nonequilibrium features in three heat-reservoir
set ups \cite{Diaz2021}. A 3LS may be modelled as
a superconducting loop containing three Josephson junctions
\cite{George2020}, or through a V-type level structure where the two
degenerate upper levels are Raman-coupled via the lower
level  \cite{Kurizki2015}.
Ref. \cite{quan2005} considered a quasi-static cycle
allowing four possible changes in the energy gaps in the first
adiabatic step of the Otto cycle:
both gaps shrinking, one gap shrinking while the other
expands, and both gaps expanding. It was shown in the
high temperature limit that the case of both gaps
shrinking yields a positive work condition that
can be looser than that of two-level system.
More recently,  in Ref. \cite{Kurizki2015},
the power output a continuous heat engine utilizing a degenerated V-type 3LS was
reported to be doubled as compared two independent two-level systems (see also \cite{Venu2017}).
Efficiency enhancement for a quasi-static model with
a non-selective measurement protocol was reported over a two-reservoir model having a 3LS with a decoupled internal level
\cite{anka2021}.
The influence of population and coherence on the performance of a three-level engine continuously coupled to hot and cold reservoirs was studied in Ref. \cite{Pena_2021}.
Although, other studies have also been
devoted to 3LS based machines
\cite{Boukozba2007, Linden2010, Kurizki2015, Venu2017, Varinder2019, Varinder2020, Macovie_2022,Deng_2023},
a complete characterization of
the operating conditions of a quasi-static Otto cycle
seems to be lacking in literature.

In this context, we address
the following question: Given two heat reservoirs
at different temperatures, what changes in the two gaps
are compatible with specific changes in the
occupation probabilities of a 3LS?
In the simplified, semi-classical picture that we
consider, we make various assumptions. First, the
interaction between the 3LS and the heat reservoirs
is considered weak so that the hamiltonian
does not contain any such interaction term.
Second, the energy gaps are modulated by
an external field such that the level populations
in the system are preserved. The average energy thus exchanged
between the field and the system is interpreted
as work only. In this, we also neglect any energy cost
for operating the external field. Since our system
undergoes complete thermalization and there
are no coherences generated in adiabatic steps,
we are able to exclude any effects on performance
due to coherence. Despite these
simplifying assumptions, we believe our analysis is relevant
for the study of quantum thermodynamics, since a central
concern there is to ascertain the nature and extent of quantum
advantage in the performance of quantum thermal machines.

So, in our analysis, the state of a 3LS is described in terms
of two energy gaps along with relevant probability distribution.
Since different stages of an Otto cycle  lead to either an exchange of
work or heat with the surroundings,
the changes incurred in the gaps and the occupation
probabilities represent two distinct types of controls
in a quantum Otto cycle.
We ascertain the
permissible regimes of operation following
changes in the gaps vis-\`{a}-vis changes between
the hot and cold probability distributions.
As our key results,
we show that at least one of the energy
gaps in a 3LS must shrink in order to realize an Otto
engine; the engine is not feasible
if both the gaps expand during the
first quantum adiabatic stage.
Further, a definite
majorization relation between the hot and
the cold distributions of 3LS
determines the feasibility of the engine.
If the cold state majorizes the hot state,
then either one or both the gaps may shrink.
When both gaps shrink, then majorization is a sufficient
condition for the engine.
When only one gap shrinks, then majorization
is a necessary, but not a sufficient
condition for the Otto engine.
Conversely, if the machine works as an engine while
only one gap shrinks, then the majorization relation must hold.

The paper is organized as follows:
In Sec. II, we consider the QOE with a 3LS as working medium.
In Sec. III, the compatibility conditions
between probability
distributions at the two reservoirs and energy gaps
 for the feasibility of an Otto engine are derived. In Sec. IV, we discuss bounds on the efficiency of the
Otto engine and compare various configurations to identify optimal performance. It is observed
that configurations yielding better performance are those where probabilities are distributed
according to majorization relation between hot and cold
distributions. In Sec. V, we explore
special cases where one of the
gaps remains fixed or the total gap may be fixed during the cycle. In Sec. VI,
the performance of a generalized n-level working medium is
studied, followed by the conclusions of our
paper in Sec VII.

\section{Quantum Otto engine}

The notions of heat and work play a central role in thermodynamics.
Extracting heat from a hot reservoir
($Q_1 > 0$), a heat engine delivers a part of this energy as work
($W < 0$)  and dumps the unutilized heat into the cold
reservoir ($Q_2 < 0$). So, in a cyclic process,
energy conservation yields:
$Q_1 + Q_2 +W =0$.
In the quasi-static Otto cycle, each step is slow enough and
the notion of rates does not enter the formalism. Each of the compression/expansion stages involves a quantum adiabatic process, where an externally controllable parameter of the Hamiltonian can be varied. The quantum adiabatic theorem \cite{Adiabatic_Fock} ensures that this process does not cause transitions between energy levels, thus  preserving their occupation probabilities.
The remaining two steps in the cycle are the isochoric heating and cooling processes, where the system thermalizes
with the hot and cold reservoirs, respectively.
The complete cycle is described in Appendix A
and a schematic is given in Fig. 1.

The general form of the heat and work contributions can be defined using the first law of thermodynamics \cite{quan2007quantum}.
For an $n$-level system, the
heat exchanges during Stage-A and Stage-C, are respectively given by
\begin{align} \label{eqn7}
	Q_{1}&=\sum_{k=1}^n\varepsilon_{k}(P_{k}-P_{k}^{\prime}),~~~~~
	Q_{2} =\sum_{k=1}^n\varepsilon_{k}^{\prime}(P_{k}^{\prime}-P_{k}),
\end{align}
where $\varepsilon_{k} (\varepsilon_{k}^{\prime})$
is an energy level of the working medium
at the hot (cold) contact, with $P_k (P_k')$ as the
corresponding canonical occupation probability.
The net work performed by the engine in one cycle
 (${\cal W} = -W$) is given by:
\begin{equation} \label{eqn8}
	{\cal W} = Q_{1}+Q_{2} = \sum_{k=1}^{n}(\varepsilon_{k}-
	\varepsilon_{k}^{\prime})(P_{k}-P_{k}^{\prime}).
\end{equation}

${\cal W} > 0$ is called the positive work condition (PWC) of  the engine. The efficiency of the engine
is defined as $\eta = {\cal W}/Q_{1} = 1 + Q_2/Q_1$.

\begin{figure}[h]
	\centering
	{\includegraphics[width=8cm]{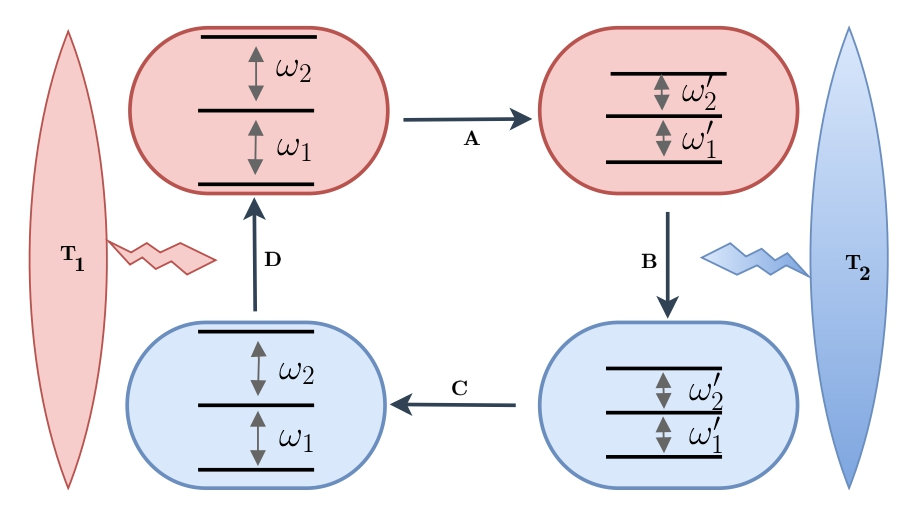}}
	\caption{The schematic of a Quantum Otto engine (QOE)
	based on a 3LS. $W_1$ and $W_2$ represent work performed during the first (A) and the second (C) quantum adiabatic
	process, respectively. Heat exchange at the hot ($T_1$) and the cold ($T_2$)  reservoir occurs in the second (B) and the fourth stage (D), denoted as $Q_1>0$ and $Q_2<0$, respectively. In the adiabatic branches, the energy gaps interchange between values $\omega_1 \leftrightarrow \omega_1'$ and
	$\omega_2 \leftrightarrow \omega_2'$.}
	\label{QOC}
\end{figure}

We consider a non-degenerate 3LS as our working medium for QOE.
The hamiltonian of the system is given by: $H = \sum_{k=1}^{3} \varepsilon_k
|k\rangle \langle k|$.
The energy eigenvalues and the
occupation probabilities at the hot and cold
 contacts  respectively are ordered as:
$\varepsilon_1 <\varepsilon_2<\varepsilon_3$
 $(P_1 > P_2 > P_3)$ and
$\varepsilon_1' <\varepsilon_2'<\varepsilon_3'$
 $(P_1' > P_2' > P_3')$.
The lower and upper energy gaps are given in terms
of transition frequencies as:
$\hbar \omega_1 =\varepsilon_2-\varepsilon_1 > 0$ and
$\hbar \omega_2=\varepsilon_3-\varepsilon_2 >0$
(see Fig. 1) with
the sum, $\omega = \omega_1+\omega_2$.
We use units where $\hbar =1$ and Boltzmann
constant $k_{\rm B} =1$.
Thus, we also define
$\omega'_1=\varepsilon'_2-\varepsilon'_1 >0$,
$\omega'_2=\varepsilon'_3-\varepsilon'_2 >0$ and
$\omega' =\omega_1'+\omega_2'$.
In terms of these gaps,
the probability distributions are written as:
$P_1 =({1+e^{-\omega_1/T_1}+e^{-\omega/T_1}})^{-1},
	P_3= ({1+e^{\omega_2/T_1}+e^{\omega/T_1}})^{-1}$ and
	$P_2 = 1-P_1 -P_3$.
	Likewise,
	$P'_1=({1+e^{-\omega'_1/T_2}+e^{-\omega'/T_2}})^{-1},
	P'_3 =({1+e^{\omega_2'/T_2}+e^{\omega'/T_2}})^{-1}$
	and
	$P'_2= 1-P'_1 - P'_3$.
Finally, the expressions for heat and work are given by
\begin{align}
Q_1 & = \omega_2(P_3-P'_3) + \omega_1(P'_1-P_1),
\label{q1}\\
	Q_2 & = -\omega'_2(P_3-P'_3) - \omega'_1(P'_1-P_1),
	\label{q2}\\
	{\cal W}
	& = (\omega_2-\omega'_2)(P_3-P'_3) + (\omega_1-\omega'_1)(P'_1-P_1).
	\label{wr3}
\end{align}
Note that the net work is the sum of the work
performed in the adiabatic stages: ${W} = W_1 + W_2$.
However, in our analysis, we only require the total work
to be extracted ($W <0$)
and do not specify the sign taken by the individual
work contributions. $W_1 <0$ and $W_2 >0$ are shown
in Fig. 1, only for concreteness.
\section{Changing the gaps vis-\`{a}-vis probabilities}
Assuming that both gaps undergo changes during the engine
cycle, the operating conditions, as
applied to Eqs. (\ref{q1})-(\ref{wr3}), imply
the following inequalities.
\begin{align}
 Q_1 > 0 & \implies 	\frac{\omega_2}{\omega_1}>\frac{P_1^{}-P_1'}{P_3^{}-P_3'},\label{}\label{}\\
 Q_2 < 0  & \implies	\frac{\omega_2'}{\omega_1'} >\frac{P_1^{}-P_1'}{P_3^{}-P_3'},\label{P9} \\
 {\cal W} > 0 & \implies	\frac{\omega_2^{}-\omega_2'}{\omega_1^{}-\omega'_1}>\frac{P_1^{}-P_1'}{P_3^{}-P_3'}\label{r5}.
\end{align}
There are four possible ways  in which the
gaps can change in the first adiabatic step.

\textbf{Cond-1} Both gaps shrink: $\omega_1 >\omega_1', ~\omega_2
> \omega_2'$.

\textbf{Cond-2} Lower gap shrinks, upper gap expands: $\omega_1
> \omega_1',~\omega_2 <\omega_2'$.

\textbf{Cond-3} Upper gap shrinks, lower gap expands: $\omega_1
<\omega_1',~\omega_2 >\omega_2'$.

\textbf{Cond-4} Both gaps expand: $\omega_1< \omega_1',~\omega_2
<\omega_2'$.

On the other hand, the changes in the occupation
probabilities may belong to one of the following cases.

\textbf{Case-(a)}	$P_1>P_1',~~P_3>P_3',~~ P_2<P_2'$

\textbf{Case-(b)}	$P_1<P_1',~~P_3<P_3',~~ P_2>P_2'$

\textbf{Case-(c)}	$P_1<P_1',~~P_3>P_3'$	

\textbf{Case-(d)}	$P_1>P_1',~~P_3<P_3'$

In cases (a) and (b), a definite  inequality
between $P_2$ and $P_2'$ holds due to the normalization
condition on probabilities.
On the other hand,
for cases (c) and (d), the inequality
between $P_2$ and $P_2'$ remains indefinite.
Next, we ascertain
the mutual compatibility between the conditions
on gaps and the feasible cases for probabilities.
\par\noindent
\subsection{\textbf{Case (a)}}
Case-(a) specifies  $P_1>P_1'$, $P_3>P_3'$ and $P_2<P_2'$.
So, the right-hand side of Eq. (\ref{r5}) is a positive quantity. Therefore, we have
\begin{align}
	\frac{\omega_2^{}-\omega_2'}{\omega_1^{}-\omega'_1}&> 0
	\label{pwca},
\end{align}
i.e. either both gaps shrink or they expand.
Using the above inequalities between the probabilities
along with their explicit forms (given in Sec. II),
we obtain
\begin{equation}\label{wr4}
	\frac{P_1}{P_2}>\frac{P_1'}{P_2'}\Rightarrow \frac{\omega_1}{T_1}>\frac{\omega_1'}{T_2}.
\end{equation}
Since $T_1 > T_2 >0$, we deduce that
$\omega_1>\omega_1'$. So,
the lower gap must shrink in the first adiabatic step.
By virtue of Eq. (\ref{pwca}),
this also implies $\omega_2>\omega_2'$.
Thus, Case (a), dictating specific inequalities between the
hot and cold probabilities, requires that both gaps must shrink
in the first adiabatic step.
We also have
\begin{equation} \label{wr41}
\frac{P_3}{P_2}>\frac{P_3'}{P_2'}\Rightarrow \frac{\omega_2}{T_1}<\frac{\omega_2'}{T_2}.
\end{equation}
Thus, Case (a) further implies that the
the upper gap is constained as follows
\begin{align}
	\frac{T_1}{T_2} \omega_2' & > \omega_2  > \omega_2'.
	\label{r7}
\end{align}
From Eqs. (\ref{wr4}) and (\ref{wr41}),
we deduce that
\begin{equation}\label{P6}
	\frac{\omega_2'}{\omega_1'} > \frac{\omega_2^{}}{\omega_1^{}}.
\end{equation}
Using (\ref{r5}), (\ref{P6}) and the mediant inequality \cite{mediant}, we can write:
\begin{equation}\label{P11}
	\frac{\omega_2'}{\omega_1'}>\frac{\omega_2^{}}{\omega_1^{}}>\frac{\omega_2^{}-\omega_2'}{\omega_1^{}-\omega_1'}>\frac{P_1^{}-P_1'}{P_3^{}-P_3'}.
\end{equation}

\subsection{\textbf{Case (b)}}
Case (b) assumes $P_1 < P_1'$, $P_3 < P_3'$ and $P_2 > P_2'$.
Proceeding along the lines of Case-(a) above, we again conclude
that both the gaps must shrink. In particular,
we obtain
\begin{align}
	\omega_2&>\omega_2',\label{r13}\\
	\frac{T_1}{T_2}\omega_1'& > \omega_1 > \omega_1'.\label{r14}
\end{align}
Then, the conditions $Q_1 >0$ and $Q_2 <0$ combined with
PWC can be expressed as:
\begin{equation}\label{P21}
	\frac{\omega_2^{}-\omega_2'}{\omega_1^{}-\omega_1'}>\frac{\omega_2^{}}{\omega_1^{}}>\frac{\omega_2'}{\omega_1'}>\frac{P_1^{}-P_1'}{P_3^{}-P_3'}>0.
\end{equation}
Hence, cases (a) and (b) are consistent
with both energy gaps shrinking during the first quantum adiabatic
process. Alternately, when we control
the 3LS as per Cond-1, then both cases (a) and (b)
are permissible.
\subsection{\textbf{Case (c)}}
This case specifies that the probabilities satisfy
$P_1<P_1'$ and $P_3>P_3'$, which
imply a majorization relation (denoted as $P \prec P'$  \cite{Marshallmajorizationbook}) between
the hot and cold distributions of a 3LS.
The said relation is usually expressed as the following
set of inequalities:
$	P_3^{} \geq P_3'$,
$P_2^{} + P_3^{}  \geq  P_2^{'} + P_3'$,
and $P_1^{} + P_2^{} + P_3 = P_1'+ P_2^{'} + P_3'=1$.
Clearly, these conditions also imply $P_1 \leq P_1'$.
Thus, for Case (c), we obtain
\begin{equation}\label{k3}
	\frac{P_3}{P_1}>\frac{P_3'}{P_1'}\Rightarrow \frac{\omega}{T_1}<\frac{\omega'}{T_2}.
\end{equation}
The extreme right-hand side of Eq. (\ref{r5}) is negative in this case. Therefore, we conclude that positive work is permissible when both gaps shrink (Cond 1), but also when only one gap shrinks while the other one expands (Cond 2 and 3).  Consequently, when the hot and cold probabilities satisfy the majorization relation, then all three conditions (Cond-1,2,3) on the energy gaps are feasible. In this sense,
 the validity of a majorization relation gives
us the maximal freedom to manipulate a 3LS for
the purpose of Otto engine.

\begin{figure}[h!]
	             \centering
				{\includegraphics[width=8cm]{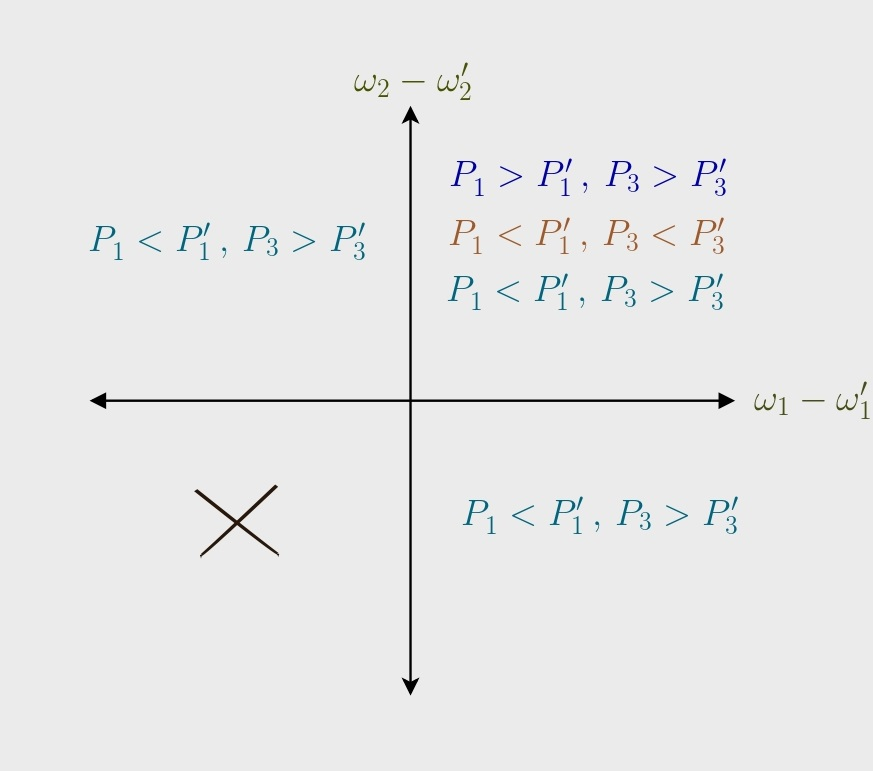}}
				\caption{In the first quadrant where both gaps  shrink, cases (a), (b), and (c) are all permissible. However, in the second and fourth quadrants, only the majorization relation
				($P \prec P'$) i.e. Case (c) is allowed. In the third quadrant, where both gaps are expanding, work cannot be extracted.}
				\label{fig_2}
\end{figure}

Finally, case-(d) results in $Q_1<0$ and $Q_2>0$, which
makes it inconsistent with the operation of an engine.
A graphical summary of the compatible changes in the gaps
versus the changes in probabilities is given in Fig. (\ref{fig_2}).

\begin{figure}[h]
	\centering
	{\includegraphics[width=7.5cm]{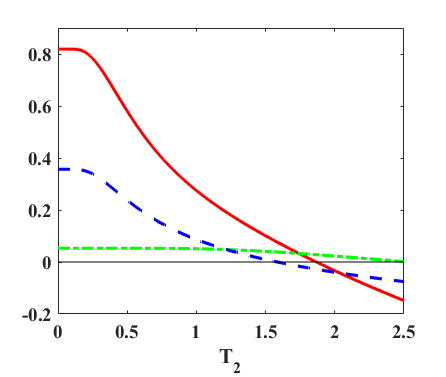}}
	\caption{
	The solid curve (red online) represents the work per cycle
	vs the temperature $T_2$ of the cold reservoir.
	The parameters are set at $T_1=4$, $J=2$, $B_1=5$ and $B_2=3$.
    With $B_1 >B_2>0$ and $J$ fixed, both gaps are shrinking.
	The dashed (blue)  and dot-dashed (green) curves
	denote  $(P_1'-P_1)$ and $(P_3 - P_3')$ respectively.
	The majorization relation ($P \prec P'$) holds
	if {\it both} dashed and dot-dashed curves are positive.
	It is seen that majorization is sufficient,
	but not a necessary condition for the work to be positive.
	}
	\label{fig_3}
\end{figure}

To illustrate the analytic observations,  we employ
a toy hamiltonian with the energy spectrum at the hot and cold thermalisation stages as ($\varepsilon_3=B_1, \varepsilon_2=-J, \varepsilon_1=-B_1)$ and $(\varepsilon'_3=B_2, \varepsilon'_2=-J, \varepsilon'_1=-B_2)$ respectively.
With $B_1 >B_2>0$ and a fixed $J$ value, it allows us to
shrink both the gaps in the first adiabatic stage.
In  Fig. \ref{fig_3}, we observe that
work may be positive whether majorization
holds or not. Thus, majorization is not a necessary
condition for PWC when both gaps are shrinking.
However, if the majorization holds ($P \prec P'$),
 then work is positive, thus making majorization
relation  a sufficient condition for Otto engine.

\begin{figure}[h]
		\centering
		{\includegraphics[width=7.5cm]{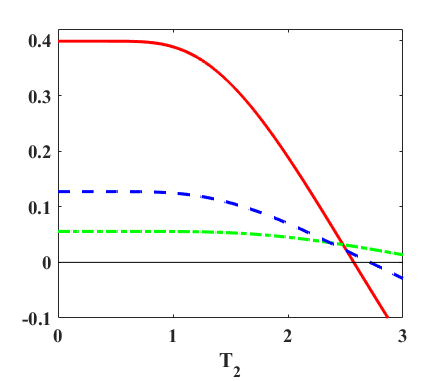}}
		\caption{Same quantities as shown in Fig. \ref{fig_3}, for the energy spectrum as [($\varepsilon_3=J, \varepsilon_2=B_1, \varepsilon_1=-B_1), ~(\varepsilon'_3=J, \varepsilon'_2=B_2, \varepsilon'_1=-B_2$)].
		The parameters are set at $T_1=4, J=6, B_1=5, B_2=3$.
		With $B_1 >B_2>0$ and $J$ fixed,
		the lower gap shrinks while the upper one expands.
		The work (solid, red) is positive
		only if the majorization condition ($P \prec P'$) is  satisfied. If
		majorization does not hold, then work is negative.}
		\label{fig_4}
\end{figure}

Similarly, in Fig. \ref{fig_4},
the energy spectrum is chosen as ($\varepsilon_3=J, \varepsilon_2=B_1, \varepsilon_1=-B_1$) and ($\varepsilon'_3=J, \varepsilon'_2=B_2, \varepsilon'_1=-B_2$). With $B_1 >B_2>0$ and fixed $J$,
		the lower gap shrinks while the upper one expands.
As we have demonstrated earlier, majorization is necessary, but not a sufficient condition for positive work extraction in this scenario.

To summarize our results so far,
if both energy gaps shrink, then
any of the  cases (a), (b) and (c) may apply.
If only one gap shrinks and
the other expands, then the majorization relation
{\it must} hold to allow the machine to function as an engine.
Note that when both the energy gaps are expanding,
the same majorization relation always holds,
but PWC does not hold.
Therefore,  Cond-(4) is forbidden for an Otto engine.

\section {Otto Efficiency}
\label{sec_eff}
A two-level Otto engine
has but one energy gap which is modulated
during the cycle ($\omega_0 \leftrightarrow
\omega_0'$). The efficiency of this cycle
is $(1-\omega_0' / \omega_0)$, where
$\omega_0' < \omega_0$ \cite{kieu2004}.
On the other hand, a 3LS has two gaps and so offers
greater flexibility. After studying
the PWC, we look at the efficiency of the cycle,
$\eta = {{\cal W}}/{Q_1}$, which is given by
\begin{align}
\eta&=\frac{(P_3^{}-P_3')(\omega_2^{}-\omega_2^{\prime})+(P'_1-P_1^{})(\omega_1^{}-\omega'_1)}{(P_3-P'_3)\omega_2+(P'_1-P_1)\omega_1},
\end{align}
which may be rearranged in the following alternate forms.
\begin{align}
\eta & =\frac{\omega_1^{}-\omega_1'}{\omega_1}\left[ \cfrac{\frac{\omega_2^{}-\omega_2'}{\omega_1^{}-\omega_1'}-\frac{P_1^{}-P_1'}{P_3^{}-P_3'}}{\frac{\omega_2}{\omega_1}-\frac{P_1^{}-P_1'}{P_3^{}-P_3'}}\right],\label{P12} \\
\eta&=\frac{\omega_2^{}-\omega_2'}{\omega_2}\left[ \frac{\frac{P_3^{}-P_3'}{P_1^{}-P_1'}-\frac{\omega_1^{}-\omega_1'}{\omega_2^{}-\omega_2'}}{\frac{P_3^{}-P_3'}{P_1^{}-P_1'}-\frac{\omega_1}{\omega_2}}\right].\label{P122}
\end{align}
Let us consider Case (a) for which both gaps are
shrinking and denote
$\xi_i \equiv ({\omega_i^{}-\omega_i'})/{\omega_i} > 0$,
where $i=1,2$.
Due to Eq. (\ref{P11}), we have $\xi_2 < \xi_1$.
Further, the factors enclosed by the square parentheses above
are less than unity. Thus, we infer that
 for Case (a), the efficiency is bounded as:
\begin{equation}
 \eta_{\rm (a)}^{} < \xi_2 < \xi_1.
 \label{etaa}
\end{equation}
Along similar lines,  we can show that Case (b)
and Cond-(1) imply:
\begin{equation}
 \eta_{\rm (b)}^{} < \xi_1  < \xi_2.
\label{etab}
 \end{equation}
Note that for Case (b), $\xi_1  < \xi_2$ holds (see Eq. (\ref{P21})).

\begin{figure}[h]
	\centering
	{\includegraphics[width=7.5cm]{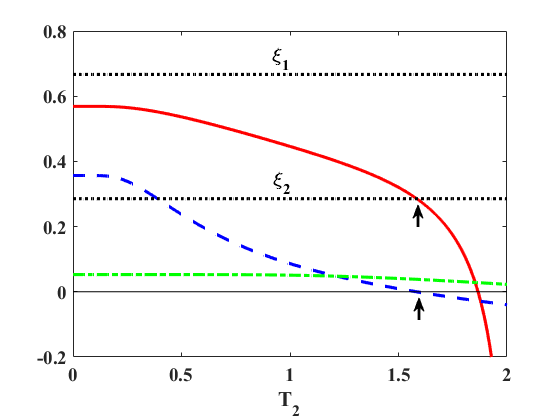}}
	\caption{
		The solid curve (red online) represents the efficiency of the Otto cycle
		while the dotted horizontal lines corresponds to $\xi_i \equiv ({\omega_i^{}-\omega_i'})/{\omega_i}$ for $i=1,2$.
		The energy spectra and the parameters are chosen as for Fig. \ref{fig_3},
		so that both gaps are shrinking and
		${\omega_2'}/{\omega_2} > {\omega_1'}/{\omega_1}$.
		As explained in the text, Case (c) applies between the bounds
		given by dotted lines (Eq. (\ref{etac})) while
		below the lower bound $\xi_2$, Case (a) is applicable (Eq. (\ref{etaa})).
		The transition from Case (c) to Case (a) happens when $P_1'-P_1$ (dotted, blue)
		changes
		sign from positive to negative (lower arrow). $P_3-P_3'$ (dot-dashed, green)
		remains positive.
		}
	\label{fig_5}
\end{figure}

Next, we consider Case-(c) where
the majorization relation holds ($P \prec P'$).
To make comparison with other cases, let us assume
that Case (a) applies, so that
${\omega_2}/{\omega_1} < {\omega_2'}/{\omega_1'}$. Then,
we can show
\begin{equation}
\xi_2 <  \eta_{\rm (c)}^{} < \xi_1.
\label{etac}
\end{equation}
Comparing with Eq. (\ref{etaa}), we conclude that majorization
or Case (c) provides a higher efficiency than Case (a).
As shown in Fig. \ref{fig_5}, within the bounds, Case (c) applies.
When the efficiency breaches the lower bound,
Case (a) is applicable.

Similarly, with both gaps shrinking and
 Case (b) as applicable, we have ${\omega_2}/{\omega_1} > {\omega_2'}/{\omega_1'}$. Then, we can derive
$\xi_2 >  \eta_{\rm (c)}^{} > \xi_1$.
Thus, again Case (c) yields a higher efficiency than
Case (b).
Therefore, we may conclude that  efficiency enhancement
can be achieved with the majorization relation when
both gaps are shrinking. On the other hand,
with one of the gaps shrinking, only the
majorization relation holds. If the lower gap
shrinks, we obtain $\eta_{\rm (c)}^{} < \xi_1$,
while if the upper gap shrinks, we have
$\eta_{\rm (c)}^{} < \xi_2$.

\section{Case of a fixed gap}
\begin{figure}[h]
		\centering
		{\includegraphics[width=7.cm]{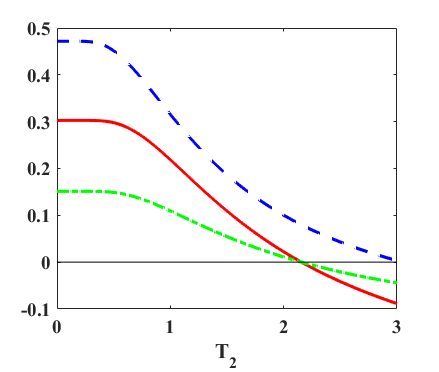}}
		\caption{The energy spectrum is chosen as ($\varepsilon_3=J_1, \varepsilon_2=B, \varepsilon_1=-B$) and ($\varepsilon'_3=J_2, \varepsilon'_2=B, \varepsilon'_1=-B$). With constant
		$B>0$ and $J_1 > J_2 >0$, the lower energy  gap is kept fixed
		while the upper gap shrinks. The parameters used are $T_1=4, B=1, J_1=4,J_2=2$.}
		\label{fig_6}
\end{figure}

The inequalities (\ref{q1})-(\ref{wr3}) are derived from the assumption that
both gaps may undergo cyclic changes during the heat cycle.
An interesting special case arises when one of the gaps
stays fixed while the other one shrinks.
The variable gap cannot expand if the machine
is to work as an engine. Now, following
the relations between probability distributions, it can
be shown that cases (a) and (b) are
not permissible when only one gap shrinks while the other stays
fixed.
Only the Case (c), satisfying the majorization condition,
allows for positive work.
Furthermore, the efficiency is bounded as:
 $\eta < \xi_i$,
where $i=1$ or 2 denoting the variable gap.
Thus, majorization provides a necessary
and sufficient condition for the
engine when one gap is held fixed,
as depicted in Figs. \ref{fig_6}.

Finally, we consider the special case where
the total gap is fixed i.e.  $\omega = \omega'$.
We then have one of the two conditions: either {Cond-{2}}
where $\omega_1>\omega_1',~~\omega_2<\omega_2'$, or
{Cond-{3}}, for which $\omega_1<\omega_1',~~ \omega_2>\omega_2'$.
The expression for work (\ref{wr3})  is now simplified to
\begin{equation}
	{\cal W} =  (\omega_2-\omega_2')(P_2-P_2') >0.
	\label{wr11}
\end{equation}
As discussed earlier, with one gap shrinking and the other gap
expanding, only Case (c) or the majorization relation holds
($P_3>P'_3,~~P_1<P_1'$).
PWC then necessitates a definite relation between $P_2$ and $P_2'$
(which is otherwise not guaranteed by the majorization condition). Thus,
Cond-{2} implies  $P_2<P_2'$, while
Cond-{3} implies $P_2>P_2'$.
Thus, along with the
majorization condition, a definite inequality between the
middle level probabilities determines
the PWC when the total gap is held fixed during the cycle.

\section{Generalization to {n}-level working medium}
Now, we look into the possible conditions for
a generic $n$-level system undergoing an Otto cycle.
From the general expressions for heat exchange, we can write
\begin{equation} \label{q1n}
	Q_{1} =\varepsilon_{1}(P_{1}^{}-P_{1}^{\prime})+\varepsilon_{2}(P_{2}^{}-P_{2}^{\prime})+\dots+\varepsilon_{n}(P_{n}^{}-P_{n}^{\prime}),
\end{equation}
and
\begin{equation} \label{}
	Q_{2} =\varepsilon_{1}^{\prime}(P_{1}^{\prime}-P_{1})+\varepsilon_{2}^{\prime}(P_{2}^{\prime}-P_2)+\dots+\varepsilon_{n}^{\prime}(P_{n}^{\prime}-P_{n}).
\end{equation}
Due to the normalization conditions,
we have $P_1^{}=1- \sum_{k=2}^{n} P_{k}^{} $ and
$P_1^{\prime}=1- \sum_{k=2}^{n} P_{k}^{'} $, which can be
substituted in Eq. (\ref{q1n}) to obtain
\begin{align} \label{q1gn}
	Q_{1}&=\varepsilon_{1}\left[ (P_n'+P_{n-1}'\dots +P_{2}^{\prime})-(P_n^{}+P_{n-1}^{}+\dots +P_2^{})\right] +\dots +\varepsilon_{n}(P_n^{}-P_{n}^{\prime}) \nonumber \\
	&=(\varepsilon_{n}-\varepsilon_{1})(P_{n}^{}-P_{n}^{\prime})+(\varepsilon_{n-1}-\varepsilon_{1})(P_{n-1}^{}-P_{n-1}^{\prime})+\dots+(\varepsilon_{2}-\varepsilon_{1})(P_{2}^{}-P_{2}^{\prime})
	\nonumber \\
	&=(\omega_1+\omega_{2}+\dots+\omega_{n-1})(P_{n}^{}-P_{n}^{\prime})+(\omega_1+\omega_2+\dots+\omega_{n-2})(P_{n-1}^{}-P_{n-1}^{\prime})+\dots \nonumber \\
	&  +\omega_1(P_{2}^{}-P_{2}^{\prime})\nonumber \\
	&=\omega_1\left[ (P_n^{}+P_{n-1}^{}+\dots+P_2^{})-(P_n'+P_{n-1}^{\prime}+\dots+P_2^{\prime})\right] \nonumber \\ & +\omega_2 \left[(P_n^{}+P_{n-1}^{}+\dots+P_3^{})-(P_n'+P_{n-1}^{\prime}+\dots+P_3)\right]+\dots+\omega_{n-1}(P_n^{}-P_n')\nonumber \\
	&=\sum_{j=1}^{n-1}\omega_j \sum_{i=j+1}^{n}(P_i^{}-P_i').
\end{align}
Similarly, we can obtain:
\begin{equation}
 Q_2 = \sum_{j=1}^{n-1}\omega_j' \sum_{i=j+1}^{n}(P_i'-P_i^{}).
 \label{q2gn}
\end{equation}
The expression for work for an $n$-level system
is then given by
\begin{align}
	{\cal W}&= \sum_{j=1}^{n-1} (\omega_j-\omega_j') \sum_{i=j+1}^{n}(P_i-P_i')
	.\label{W1}
\end{align}
Upon examining the expressions above, we observe that the coefficients of the energy gaps in Eqs. (\ref{q1gn}) and (\ref{q2gn}) take
on a definite sign if the majorization relation
$P \prec P'$ is obeyed between the thermal distributions.
Hence, for the given energy spectra, if the cold distribution majorizes the hot distribution, then the heat exchange at the hot reservoir is always positive ($Q_1 > 0$) and the heat exchange at the cold reservoir is always negative ($Q_2 < 0$). Similarly, Eq. (\ref{W1}) reveals that the coefficient of the difference
$(\omega_j-\omega_j')$ takes
on a positive sign if the above majorization relation holds.
So, additionally if some energy gaps shrink ($\omega_j-\omega_j'>0$) while others stay
fixed ($\omega_k-\omega_k' =0$), then a positive work regime follows
automatically. For PWC, at least one gap must shrink while others may be
kept fixed. In this manner, majorization serves as a sufficient criterion for a generic $n$-level working medium undergoing a quantum Otto cycle.
Examples include coupled-spins model studied in Refs. \cite{venu2021, Sachin2023}.

\section{conclusions}
We have analyzed the mutually compatible changes in the energy gaps
and the thermal distributions in a 3LS based quantum Otto cycle.
Within the scope of our model, some interesting observations
have been made. First and foremost, it is essential for at least one energy gap to shrink
in the first adiabatic stage of the given cycle to achieve the operation as an Otto engine. When both gaps expand, net work cannot be extracted. Secondly,
when both gaps shrink, the majorization condition ($P \prec P'$) serves as a sufficient criterion to obtain an engine.
Further, it proves to be  both necessary and sufficient condition
in case the gaps shrink by the same ratio.
On the other hand,
 when one gap shrinks while the other  expands, majorization becomes necessary, though not a sufficient condition for the engine operation.
Also, it is observed that for both gaps shrinking,
Otto efficiency can be enhanced
for Case (c) compared to Cases (a) and (b).
In the special case where one of the two gaps
stays fixed during the cycle,
the validity of majorization relation is the only possible scenario
in favor of an engine, making majorization a necessary and sufficient condition.
Extending the study to an $n$-level working medium, it is shown that majorization is sufficient to determine the correct direction of
heat exchange that allows for an engine.  Moreover,
if some gaps only  shrink while others stay fixed,
then positive work is obtained in the presence of majorization. Thus, we see
a decisive role played by the notion of majorization in
the operation of an Otto engine based on a 3LS.

Our results provide a benchmark to characterize the performance
of 3-level Otto engine which can be useful to make comparisons
with more sophisticated models of Otto engine, such as those
including the effects of finite time, incomplete thermalization,
coherence and so on.
It will be interesting to perform this analysis in other generic
settings, say,  when the usual
equilibrium reservoirs are replaced by squeezed or engineered
reservoirs and to determine the corresponding operating conditions.
This will help determine the scope and limitations of the
concept of majorization in these settings.
This approach
is useful when we wish to gain analytic insight
into the performance of multi-level systems
without making the problem intractable.

\begin{acknowledgments}
S.S. acknowledges financial support in the form of a Senior
Research Fellowship from the Council for Scientific and
Industrial Research (CSIR) via Award No.
09/947(0250)/2020-EMR-I, India.
\end{acknowledgments}
\section{Appendix}
\subsection{Quantum Otto cycle}
\par\noindent
The four stages of the Otto cycle are as follows:
\par\noindent
\textbf{Stage A}
Consider a quantum system with Hamiltonian
$H(B) = \sum_k \varepsilon_k(B) | k \rangle \langle k |$,
where $B$ is an externally controllable parameter such as
the strength of magnetic field.
Initially, the system is in a thermal state,
$\rho = \sum_k P_k | k \rangle \langle k |$,
corresponding to the hot
temperature $T_1$. The occupation probability
for an energy level with energy $\varepsilon_{k}$,
is given as: $P_{k}=e^{-\varepsilon_{k}/T_{1}}/
{\sum_{k} e^{-\varepsilon_{k}/T_{1}}}$.
\par\noindent
\textbf{Stage B} The system is disconnected from the hot reservoir and undergoes a quantum adiabatic process where the external field strength is {\it lowered} from $B_{1}$ to $B_{2}$,
so the energy eigenvalues become $\varepsilon_{k}^{\prime}$. This process is governed by the quantum adiabatic theorem,
and so the occupation probabilities remain unchanged.
\par\noindent
\textbf{Stage C} The system is brought into thermal contact with the cold reservoir at temperature $T_{2} (<T_{1})$ until it achieves equilibrium with it. The energy eigenvalues remain at $\varepsilon_{k}^{\prime}$
while the occupation probabilities change from $P_{k}$ to $P_{k}^{\prime}=e^{-\varepsilon_{k}^{\prime}/T_{2}}
/{\sum_{k} e^{-\varepsilon_{k}^{\prime}/T_{2}}}$.
\par\noindent
\textbf{Stage D} The system is disengaged from the cold reservoir, and the field strength is reverted to $B_{1}$. The occupation probabilities ${P_{k}^{\prime}}$ remain unchanged, while the energy levels return to ${\varepsilon_{k}}$.

Subsequently, the system is reconnected to the hot reservoir, restoring the initial state ($\rho$) and completing one heat cycle. During an isochoric process, only heat is exchanged between the system and the reservoir, given by the difference between the final and initial mean energies of the system during that process. On the other hand, the adiabatic branches of the quantum Otto cycle solely involve work.

\subsection{When one gap is kept fixed}
We determine the PWC when the lower energy gap remains fixed
 ($\omega_1=\omega'_1$) during the Otto cycle.
The expressions for work and heat are simplified to:
\begin{align}
	{\cal W} & =(\omega_2-\omega'_2)(P_3-P'_3),\label{T4}\\
	Q_1&= \omega_2(P_3-P'_3) + \omega_1(P'_1-P_1),\\
	Q_2&= -\omega'_2(P_3-P'_3) - \omega'_1(P'_1-P_1).
\end{align}
There are four possible relationships between the
hot and cold distributions.
\par\noindent
Case-(a) yields
\begin{equation}\label{1}
	\frac{P_3}{P_2}>\frac{P'_3}{P'_2} \Rightarrow \frac{\omega_2}{T_1}<\frac{\omega'_2}{T_2}.
\end{equation}
Also, using the fixed lower gap condition, we obtain
\begin{equation}\label{2}
	\frac{P_1}{P_2}>\frac{P'_1}{P'_2}\Rightarrow \frac{1}{T_1}>\frac{1}{T_2}.
\end{equation}
But, PWC from Eq. (\ref{T4}) requires $\omega_2>\omega_2'$.
Since we have assumed $T_1>T_2$, we find Eq. (\ref{2}) in contradiction. Consequently, Case-(a) is not permissible
as Otto engine when the lower gap is fixed and the
upper gap shrinks.

Similarly, for Case-(b),
${\cal W>}0$ implies $\omega_2<\omega_2'$, since $P_3 < P_3'$.
Then, we have
\begin{equation}\label{3}
	\frac{P_3}{P_2}<\frac{P'_3}{P'_2}\Rightarrow \frac{\omega_2}{T_1}>\frac{\omega'_2}{T_2},
\end{equation}
and
\begin{equation}\label{4}
	\frac{P_1}{P_2}<\frac{P'_1}{P'_2}\Rightarrow \frac{1}{T_1}<\frac{1}{T_2}.
\end{equation}
Since, we require   $\omega_2<\omega_2'$,
Eq. (\ref{3}) cannot hold. Thus Case-(b) does not allow
for the engine.

Now, consider Case (c), which yields
\begin{equation}\label{5}
	\frac{P_3}{P_1}>\frac{P'_3}{P'_1}\Rightarrow \frac{\omega}{T_1}<\frac{\omega'}{T_2},
\end{equation}
where $\omega (\omega')$ represent the sum of the gaps
under respective conditions.
PWC from Eq. (\ref{T4}) requires $\omega_2 > \omega_2'$.
Therefore, we must have $\omega > \omega'$, as
the other gap is fixed.
Thus,  Eq. (\ref{5}) holds
as well as $Q_1>0,~~ Q_2<0$ for Case (c).
Finally, for Case-(d), we obtain
$Q_1<0,~~ Q_2>0$. Thus, engine is not
a permissible operation with Case (d).
Therefore, only the majorization condition
permits the operation as an engine.

Similarly, we can treat the case of the fixed upper gap ($\omega_2=\omega'_2$)
while the lower gap is shrinking. Here also, it can be shown that only
the majorization condition allows for the engine operation.

%


\begin{thebibliography}{121}%
\makeatletter
\providecommand \@ifxundefined [1]{%
 \@ifx{#1\undefined}
}%
\providecommand \@ifnum [1]{%
 \ifnum #1\expandafter \@firstoftwo
 \else \expandafter \@secondoftwo
 \fi
}%
\providecommand \@ifx [1]{%
 \ifx #1\expandafter \@firstoftwo
 \else \expandafter \@secondoftwo
 \fi
}%
\providecommand \natexlab [1]{#1}%
\providecommand \enquote  [1]{``#1''}%
\providecommand \bibnamefont  [1]{#1}%
\providecommand \bibfnamefont [1]{#1}%
\providecommand \citenamefont [1]{#1}%
\providecommand \href@noop [0]{\@secondoftwo}%
\providecommand \href [0]{\begingroup \@sanitize@url \@href}%
\providecommand \@href[1]{\@@startlink{#1}\@@href}%
\providecommand \@@href[1]{\endgroup#1\@@endlink}%
\providecommand \@sanitize@url [0]{\catcode `\\12\catcode `\$12\catcode
  `\&12\catcode `\#12\catcode `\^12\catcode `\_12\catcode `\%12\relax}%
\providecommand \@@startlink[1]{}%
\providecommand \@@endlink[0]{}%
\providecommand \url  [0]{\begingroup\@sanitize@url \@url }%
\providecommand \@url [1]{\endgroup\@href {#1}{\urlprefix }}%
\providecommand \urlprefix  [0]{URL }%
\providecommand \Eprint [0]{\href }%
\providecommand \doibase [0]{https://doi.org/}%
\providecommand \selectlanguage [0]{\@gobble}%
\providecommand \bibinfo  [0]{\@secondoftwo}%
\providecommand \bibfield  [0]{\@secondoftwo}%
\providecommand \translation [1]{[#1]}%
\providecommand \BibitemOpen [0]{}%
\providecommand \bibitemStop [0]{}%
\providecommand \bibitemNoStop [0]{.\EOS\space}%
\providecommand \EOS [0]{\spacefactor3000\relax}%
\providecommand \BibitemShut  [1]{\csname bibitem#1\endcsname}%
\let\auto@bib@innerbib\@empty
\bibitem [{\citenamefont {Scovil}\ and\ \citenamefont
  {Schulz-DuBois}(1959)}]{Threelevelmaser}%
  \BibitemOpen
  \bibfield  {author} {\bibinfo {author} {\bibfnamefont {H.~E.~D.}\
  \bibnamefont {Scovil}}\ and\ \bibinfo {author} {\bibfnamefont {E.~O.}\
  \bibnamefont {Schulz-DuBois}},\ }\bibfield  {title} {\bibinfo {title}
  {Three-level masers as heat engines},\ }\href
  {https://doi.org/10.1103/PhysRevLett.2.262} {\bibfield  {journal} {\bibinfo
  {journal} {Phys. Rev. Lett.}\ }\textbf {\bibinfo {volume} {2}},\ \bibinfo
  {pages} {262} (\bibinfo {year} {1959})}\BibitemShut {NoStop}%
\bibitem [{\citenamefont {Alicki}(1979)}]{Alicki1979}%
  \BibitemOpen
  \bibfield  {author} {\bibinfo {author} {\bibfnamefont {R.}~\bibnamefont
  {Alicki}},\ }\bibfield  {title} {\bibinfo {title} {The quantum open system as
  a model of the heat engine},\ }\href
  {https://doi.org/10.1088/0305-4470/12/5/007} {\bibfield  {journal} {\bibinfo
  {journal} {Journal of Physics A: Mathematical and General}\ }\textbf
  {\bibinfo {volume} {12}},\ \bibinfo {pages} {L103} (\bibinfo {year}
  {1979})}\BibitemShut {NoStop}%
\bibitem [{\citenamefont {Lindblad}(1976)}]{linblad1976}%
  \BibitemOpen
  \bibfield  {author} {\bibinfo {author} {\bibfnamefont {G.}~\bibnamefont
  {Lindblad}},\ }\bibfield  {title} {\bibinfo {title} {{On the generators of
  quantum dynamical semigroups}},\ }\href
  {https://doi.org/https://doi.org/10.1007/BF01608499} {\bibfield  {journal}
  {\bibinfo  {journal} {Communications in Mathematical Physics}\ }\textbf
  {\bibinfo {volume} {48}},\ \bibinfo {pages} {119 } (\bibinfo {year}
  {1976})}\BibitemShut {NoStop}%
\bibitem [{\citenamefont {Sheehan}(2002)}]{Sheehan2002}%
  \BibitemOpen
  \bibfield  {author} {\bibinfo {author} {\bibfnamefont {D.~P.}\ \bibnamefont
  {Sheehan}},\ }\bibfield  {title} {\bibinfo {title} {Quantum limits to the
  {S}econd law: {F}irst {I}nternational conference on quantum limits to the
  {S}econd law, {S}an {D}iego, {C}alifornia 28-31 {J}uly 2002}\ }(\bibinfo
  {year} {2002})\BibitemShut {NoStop}%
  \bibitem [{\citenamefont {Scully}\ \emph {et~al.}(2003)\citenamefont {Scully},
  \citenamefont {Zubairy}, \citenamefont {Agarwal},\ and\ \citenamefont
  {Walther}}]{Scully2003}%
  \BibitemOpen
  \bibfield  {author} {\bibinfo {author} {\bibfnamefont {M.~O.}\ \bibnamefont
  {Scully}}, \bibinfo {author} {\bibfnamefont {M.~S.}\ \bibnamefont {Zubairy}},
  \bibinfo {author} {\bibfnamefont {G.~S.}\ \bibnamefont {Agarwal}},\ and\
  \bibinfo {author} {\bibfnamefont {H.}~\bibnamefont {Walther}},\ }\bibfield
  {title} {\bibinfo {title} {Extracting work from a single heat bath via
  vanishing quantum coherence},\ }\href
  {https://doi.org/10.1126/science.1078955} {\bibfield  {journal} {\bibinfo
  {journal} {Science}\ }\textbf {\bibinfo {volume} {299}},\ \bibinfo {pages}
  {862} (\bibinfo {year} {2003})}\BibitemShut {NoStop}%
\bibitem [{\citenamefont {Allahverdyan}\ \emph {et~al.}(2004)\citenamefont
  {Allahverdyan}, \citenamefont {Balian},\ and\ \citenamefont
  {Nieuwenhuizen}}]{Allahverdyan2004}%
  \BibitemOpen
  \bibfield  {author} {\bibinfo {author} {\bibfnamefont {A.~E.}\ \bibnamefont
  {Allahverdyan}}, \bibinfo {author} {\bibfnamefont {R.}~\bibnamefont
  {Balian}},\ and\ \bibinfo {author} {\bibfnamefont {T.~M.}\ \bibnamefont
  {Nieuwenhuizen}},\ }\bibfield  {title} {\bibinfo {title} {Maximal work
  extraction from finite quantum systems},\ }\href
  {https://doi.org/10.1209/epl/i2004-10101-2} {\bibfield  {journal} {\bibinfo
  {journal} {Europhysics Letters (EPL)}\ }\textbf {\bibinfo {volume} {67}},\
  \bibinfo {pages} {565–571} (\bibinfo {year} {2004})}\BibitemShut {NoStop}%
\bibitem [{\citenamefont {Gemmer}\ \emph {et~al.}(2009)\citenamefont {Gemmer},
  \citenamefont {Michel},\ and\ \citenamefont {Mahler}}]{Mahlerbook1}%
  \BibitemOpen
  \bibfield  {author} {\bibinfo {author} {\bibfnamefont {J.}~\bibnamefont
  {Gemmer}}, \bibinfo {author} {\bibfnamefont {M.}~\bibnamefont {Michel}},\
  and\ \bibinfo {author} {\bibfnamefont {G.}~\bibnamefont {Mahler}},\ }\href
  {https://doi.org/https://doi.org/10.1007/978-3-540-70510-9} {\emph {\bibinfo
  {title} {Quantum Thermodynamics: Emergence of Thermodynamic Behavior Within
  Composite Quantum System}}},\ \bibinfo {edition} {2nd}\ ed.,\ Lecture Notes
  in Physics\ (\bibinfo  {publisher} {Springer Berlin, Heidelberg},\ \bibinfo
  {year} {2009})\BibitemShut {NoStop}%
\bibitem [{\citenamefont {Allahverdyan}\ \emph {et~al.}(2008)\citenamefont
  {Allahverdyan}, \citenamefont {Johal},\ and\ \citenamefont
  {Mahler}}]{Allahverdyan2008}%
  \BibitemOpen
  \bibfield  {author} {\bibinfo {author} {\bibfnamefont {A.~E.}\ \bibnamefont
  {Allahverdyan}}, \bibinfo {author} {\bibfnamefont {R.~S.}\ \bibnamefont
  {Johal}},\ and\ \bibinfo {author} {\bibfnamefont {G.}~\bibnamefont
  {Mahler}},\ }\bibfield  {title} {\bibinfo {title} {Work extremum principle:
  Structure and function of quantum heat engines},\ }\href
  {https://doi.org/10.1103/PhysRevE.77.041118} {\bibfield  {journal} {\bibinfo
  {journal} {Phys. Rev. E}\ }\textbf {\bibinfo {volume} {77}},\ \bibinfo
  {pages} {041118} (\bibinfo {year} {2008})}\BibitemShut {NoStop}%
\bibitem [{\citenamefont {Quan}\ \emph {et~al.}(2007)\citenamefont {Quan},
  \citenamefont {Liu}, \citenamefont {Sun},\ and\ \citenamefont
  {Nori}}]{quan2007quantum}%
  \BibitemOpen
  \bibfield  {author} {\bibinfo {author} {\bibfnamefont {H.}~\bibnamefont
  {Quan}}, \bibinfo {author} {\bibfnamefont {Y.-X.}\ \bibnamefont {Liu}},
  \bibinfo {author} {\bibfnamefont {C.}~\bibnamefont {Sun}},\ and\ \bibinfo
  {author} {\bibfnamefont {F.}~\bibnamefont {Nori}},\ }\bibfield  {title}
  {\bibinfo {title} {Quantum thermodynamic cycles and quantum heat engines},\
  }\href {https://doi.org/10.1103/PhysRevE.76.031105} {\bibfield  {journal}
  {\bibinfo  {journal} {Phys. Rev. E}\ }\textbf {\bibinfo {volume} {76}},\
  \bibinfo {pages} {031105} (\bibinfo {year} {2007})}\BibitemShut {NoStop}%
\bibitem [{\citenamefont {Myers}\ \emph {et~al.}(2022)\citenamefont {Myers},
  \citenamefont {Abah},\ and\ \citenamefont {Deffner}}]{Myers_2022}%
  \BibitemOpen
  \bibfield  {author} {\bibinfo {author} {\bibfnamefont {N.~M.}\ \bibnamefont
  {Myers}}, \bibinfo {author} {\bibfnamefont {O.}~\bibnamefont {Abah}},\ and\
  \bibinfo {author} {\bibfnamefont {S.}~\bibnamefont {Deffner}},\ }\bibfield
  {title} {\bibinfo {title} {Quantum thermodynamic devices: From theoretical
  proposals to experimental reality},\ }\href
  {https://doi.org/https://doi.org/10.1116/5.0083192} {\bibfield  {journal}
  {\bibinfo  {journal} {{AVS} Quantum Science}\ }\textbf {\bibinfo {volume}
  {4}},\ \bibinfo {pages} {027101} (\bibinfo {year} {2022})}\BibitemShut
  {NoStop}%
\bibitem [{\citenamefont {Vinjanampathy}\ and\ \citenamefont
  {Anders}(2016)}]{Vinjanampathy_2016}%
  \BibitemOpen
  \bibfield  {author} {\bibinfo {author} {\bibfnamefont {S.}~\bibnamefont
  {Vinjanampathy}}\ and\ \bibinfo {author} {\bibfnamefont {J.}~\bibnamefont
  {Anders}},\ }\bibfield  {title} {\bibinfo {title} {Quantum thermodynamics},\
  }\href {https://doi.org/10.1080/00107514.2016.1201896} {\bibfield  {journal}
  {\bibinfo  {journal} {Contemporary Physics}\ }\textbf {\bibinfo {volume}
  {57}},\ \bibinfo {pages} {545} (\bibinfo {year} {2016})}\BibitemShut
  {NoStop}%
\bibitem [{\citenamefont {Millen}\ and\ \citenamefont
  {Xuereb}(2016)}]{Millen_2016}%
  \BibitemOpen
  \bibfield  {author} {\bibinfo {author} {\bibfnamefont {J.}~\bibnamefont
  {Millen}}\ and\ \bibinfo {author} {\bibfnamefont {A.}~\bibnamefont
  {Xuereb}},\ }\bibfield  {title} {\bibinfo {title} {Perspective on quantum
  thermodynamics},\ }\href {https://doi.org/10.1088/1367-2630/18/1/011002}
  {\bibfield  {journal} {\bibinfo  {journal} {New Journal of Physics}\ }\textbf
  {\bibinfo {volume} {18}},\ \bibinfo {pages} {011002} (\bibinfo {year}
  {2016})}\BibitemShut {NoStop}%
\bibitem [{\citenamefont {Quan}(2009)}]{quan2009}%
  \BibitemOpen
  \bibfield  {author} {\bibinfo {author} {\bibfnamefont {H.~T.}\ \bibnamefont
  {Quan}},\ }\bibfield  {title} {\bibinfo {title} {Quantum thermodynamic cycles
  and quantum heat engines. ii.},\ }\href
  {https://doi.org/10.1103/PhysRevE.79.041129} {\bibfield  {journal} {\bibinfo
  {journal} {Phys. Rev. E}\ }\textbf {\bibinfo {volume} {79}},\ \bibinfo
  {pages} {041129} (\bibinfo {year} {2009})}\BibitemShut {NoStop}%
\bibitem [{\citenamefont {Altintas}\ \emph {et~al.}(2014)\citenamefont
  {Altintas}, \citenamefont {Hardal},\ and\ \citenamefont {M\"ustecapl\ifmmode
  \imath \else \i \fi{}o\ifmmode~\breve{g}\else \u{g}\fi{}lu}}]{altintas2014}%
  \BibitemOpen
  \bibfield  {author} {\bibinfo {author} {\bibfnamefont {F.}~\bibnamefont
  {Altintas}}, \bibinfo {author} {\bibfnamefont {A.~U.~C.}\ \bibnamefont
  {Hardal}},\ and\ \bibinfo {author} {\bibfnamefont {O.~E.}\ \bibnamefont
  {M\"ustecapl\ifmmode \imath \else \i \fi{}o\ifmmode~\breve{g}\else
  \u{g}\fi{}lu}},\ }\bibfield  {title} {\bibinfo {title} {Quantum correlated
  heat engine with spin squeezing},\ }\href
  {https://doi.org/10.1103/PhysRevE.90.032102} {\bibfield  {journal} {\bibinfo
  {journal} {Phys. Rev. E}\ }\textbf {\bibinfo {volume} {90}},\ \bibinfo
  {pages} {032102} (\bibinfo {year} {2014})}\BibitemShut {NoStop}%
\bibitem [{\citenamefont {Barrios}\ \emph {et~al.}(2017)\citenamefont
  {Barrios}, \citenamefont {Albarr\'an-Arriagada}, \citenamefont
  {C\'ardenas-L\'opez}, \citenamefont {Romero},\ and\ \citenamefont
  {Retamal}}]{borris2017}%
  \BibitemOpen
  \bibfield  {author} {\bibinfo {author} {\bibfnamefont {G.~A.}\ \bibnamefont
  {Barrios}}, \bibinfo {author} {\bibfnamefont {F.}~\bibnamefont
  {Albarr\'an-Arriagada}}, \bibinfo {author} {\bibfnamefont {F.~A.}\
  \bibnamefont {C\'ardenas-L\'opez}}, \bibinfo {author} {\bibfnamefont
  {G.}~\bibnamefont {Romero}},\ and\ \bibinfo {author} {\bibfnamefont {J.~C.}\
  \bibnamefont {Retamal}},\ }\bibfield  {title} {\bibinfo {title} {Role of
  quantum correlations in light-matter quantum heat engines},\ }\href
  {https://doi.org/10.1103/PhysRevA.96.052119} {\bibfield  {journal} {\bibinfo
  {journal} {Phys. Rev. A}\ }\textbf {\bibinfo {volume} {96}},\ \bibinfo
  {pages} {052119} (\bibinfo {year} {2017})}\BibitemShut {NoStop}%
\bibitem [{\citenamefont {Um}\ \emph {et~al.}(2022)\citenamefont {Um},
  \citenamefont {Dorfman},\ and\ \citenamefont {Park}}]{Jaegon2022}%
  \BibitemOpen
  \bibfield  {author} {\bibinfo {author} {\bibfnamefont {J.}~\bibnamefont
  {Um}}, \bibinfo {author} {\bibfnamefont {K.~E.}\ \bibnamefont {Dorfman}},\
  and\ \bibinfo {author} {\bibfnamefont {H.}~\bibnamefont {Park}},\ }\bibfield
  {title} {\bibinfo {title} {Coherence-enhanced quantum-dot heat engine},\
  }\href {https://doi.org/10.1103/PhysRevResearch.4.L032034} {\bibfield
  {journal} {\bibinfo  {journal} {Phys. Rev. Res.}\ }\textbf {\bibinfo {volume}
  {4}},\ \bibinfo {pages} {L032034} (\bibinfo {year} {2022})}\BibitemShut
  {NoStop}%
\bibitem [{\citenamefont {Camati}\ \emph {et~al.}(2019)\citenamefont {Camati},
  \citenamefont {Santos},\ and\ \citenamefont {Serra}}]{Camati2019}%
  \BibitemOpen
  \bibfield  {author} {\bibinfo {author} {\bibfnamefont {P.~A.}\ \bibnamefont
  {Camati}}, \bibinfo {author} {\bibfnamefont {J.~F.~G.}\ \bibnamefont
  {Santos}},\ and\ \bibinfo {author} {\bibfnamefont {R.~M.}\ \bibnamefont
  {Serra}},\ }\bibfield  {title} {\bibinfo {title} {Coherence effects in the
  performance of the quantum {O}tto heat engine},\ }\href
  {https://doi.org/10.1103/PhysRevA.99.062103} {\bibfield  {journal} {\bibinfo
  {journal} {Phys. Rev. A}\ }\textbf {\bibinfo {volume} {99}},\ \bibinfo
  {pages} {062103} (\bibinfo {year} {2019})}\BibitemShut {NoStop}%
\bibitem [{\citenamefont {Zhang}\ \emph {et~al.}(2007)\citenamefont {Zhang},
  \citenamefont {Liu}, \citenamefont {Chen},\ and\ \citenamefont
  {Li}}]{zhang2007four}%
  \BibitemOpen
  \bibfield  {author} {\bibinfo {author} {\bibfnamefont {T.}~\bibnamefont
  {Zhang}}, \bibinfo {author} {\bibfnamefont {W.-T.}\ \bibnamefont {Liu}},
  \bibinfo {author} {\bibfnamefont {P.-X.}\ \bibnamefont {Chen}},\ and\
  \bibinfo {author} {\bibfnamefont {C.-Z.}\ \bibnamefont {Li}},\ }\bibfield
  {title} {\bibinfo {title} {Four-level entangled quantum heat engines},\
  }\href {https://doi.org/https://link.aps.org/doi/10.1103/PhysRevA.75.062102}
  {\bibfield  {journal} {\bibinfo  {journal} {Phys. Rev. A}\ }\textbf {\bibinfo
  {volume} {75}},\ \bibinfo {pages} {062102} (\bibinfo {year}
  {2007})}\BibitemShut {NoStop}%
\bibitem [{\citenamefont {Niedenzu}\ \emph {et~al.}(2015)\citenamefont
  {Niedenzu}, \citenamefont {Gelbwaser-Klimovsky},\ and\ \citenamefont
  {Kurizki}}]{Niedenzu2015}%
  \BibitemOpen
  \bibfield  {author} {\bibinfo {author} {\bibfnamefont {W.}~\bibnamefont
  {Niedenzu}}, \bibinfo {author} {\bibfnamefont {D.}~\bibnamefont
  {Gelbwaser-Klimovsky}},\ and\ \bibinfo {author} {\bibfnamefont
  {G.}~\bibnamefont {Kurizki}},\ }\bibfield  {title} {\bibinfo {title}
  {Performance limits of multilevel and multipartite quantum heat machines},\
  }\href {https://doi.org/10.1103/PhysRevE.92.042123} {\bibfield  {journal}
  {\bibinfo  {journal} {Phys. Rev. E}\ }\textbf {\bibinfo {volume} {92}},\
  \bibinfo {pages} {042123} (\bibinfo {year} {2015})}\BibitemShut {NoStop}%
\bibitem [{\citenamefont {Kamimura}\ \emph {et~al.}(2022)\citenamefont
  {Kamimura}, \citenamefont {Hakoshima}, \citenamefont {Matsuzaki},
  \citenamefont {Yoshida},\ and\ \citenamefont {Tokura}}]{Kamimura2022}%
  \BibitemOpen
  \bibfield  {author} {\bibinfo {author} {\bibfnamefont {S.}~\bibnamefont
  {Kamimura}}, \bibinfo {author} {\bibfnamefont {H.}~\bibnamefont {Hakoshima}},
  \bibinfo {author} {\bibfnamefont {Y.}~\bibnamefont {Matsuzaki}}, \bibinfo
  {author} {\bibfnamefont {K.}~\bibnamefont {Yoshida}},\ and\ \bibinfo {author}
  {\bibfnamefont {Y.}~\bibnamefont {Tokura}},\ }\bibfield  {title} {\bibinfo
  {title} {Quantum-enhanced heat engine based on superabsorption},\ }\href
  {https://doi.org/10.1103/PhysRevLett.128.180602} {\bibfield  {journal}
  {\bibinfo  {journal} {Phys. Rev. Lett.}\ }\textbf {\bibinfo {volume} {128}},\
  \bibinfo {pages} {180602} (\bibinfo {year} {2022})}\BibitemShut {NoStop}%
\bibitem [{\citenamefont {Peterson}\ \emph {et~al.}(2019)\citenamefont
  {Peterson}, \citenamefont {Batalh\~ao}, \citenamefont {Herrera},
  \citenamefont {Souza}, \citenamefont {Sarthour}, \citenamefont {Oliveira},\
  and\ \citenamefont {Serra}}]{peterson2019experimental}%
  \BibitemOpen
  \bibfield  {author} {\bibinfo {author} {\bibfnamefont {J.~P.~S.}\
  \bibnamefont {Peterson}}, \bibinfo {author} {\bibfnamefont {T.~B.}\
  \bibnamefont {Batalh\~ao}}, \bibinfo {author} {\bibfnamefont
  {M.}~\bibnamefont {Herrera}}, \bibinfo {author} {\bibfnamefont {A.~M.}\
  \bibnamefont {Souza}}, \bibinfo {author} {\bibfnamefont {R.~S.}\ \bibnamefont
  {Sarthour}}, \bibinfo {author} {\bibfnamefont {I.~S.}\ \bibnamefont
  {Oliveira}},\ and\ \bibinfo {author} {\bibfnamefont {R.~M.}\ \bibnamefont
  {Serra}},\ }\bibfield  {title} {\bibinfo {title} {Experimental
  characterization of a spin quantum heat engine},\ }\href
  {https://doi.org/10.1103/PhysRevLett.123.240601} {\bibfield  {journal}
  {\bibinfo  {journal} {Phys. Rev. Lett.}\ }\textbf {\bibinfo {volume} {123}},\
  \bibinfo {pages} {240601} (\bibinfo {year} {2019})}\BibitemShut {NoStop}%
\bibitem [{\citenamefont {de~Assis}\ \emph {et~al.}(2019)\citenamefont
  {de~Assis}, \citenamefont {de~Mendon\ifmmode~\mbox{\c{c}}\else \c{c}\fi{}a},
  \citenamefont {Villas-Boas}, \citenamefont {de~Souza}, \citenamefont
  {Sarthour}, \citenamefont {Oliveira},\ and\ \citenamefont
  {de~Almeida}}]{assis2019}%
  \BibitemOpen
  \bibfield  {author} {\bibinfo {author} {\bibfnamefont {R.~J.}\ \bibnamefont
  {de~Assis}}, \bibinfo {author} {\bibfnamefont {T.~M.}\ \bibnamefont
  {de~Mendon\ifmmode~\mbox{\c{c}}\else \c{c}\fi{}a}}, \bibinfo {author}
  {\bibfnamefont {C.~J.}\ \bibnamefont {Villas-Boas}}, \bibinfo {author}
  {\bibfnamefont {A.~M.}\ \bibnamefont {de~Souza}}, \bibinfo {author}
  {\bibfnamefont {R.~S.}\ \bibnamefont {Sarthour}}, \bibinfo {author}
  {\bibfnamefont {I.~S.}\ \bibnamefont {Oliveira}},\ and\ \bibinfo {author}
  {\bibfnamefont {N.~G.}\ \bibnamefont {de~Almeida}},\ }\bibfield  {title}
  {\bibinfo {title} {Efficiency of a quantum {O}tto heat engine operating under
  a reservoir at effective negative temperatures},\ }\href
  {https://doi.org/10.1103/PhysRevLett.122.240602} {\bibfield  {journal}
  {\bibinfo  {journal} {Phys. Rev. Lett.}\ }\textbf {\bibinfo {volume} {122}},\
  \bibinfo {pages} {240602} (\bibinfo {year} {2019})}\BibitemShut {NoStop}%
\bibitem [{\citenamefont {Nettersheim}\ \emph {et~al.}(2022)\citenamefont
  {Nettersheim}, \citenamefont {Burgardt}, \citenamefont {Bouton},
  \citenamefont {Adam}, \citenamefont {Lutz},\ and\ \citenamefont
  {Widera}}]{netterahein2022}%
  \BibitemOpen
  \bibfield  {author} {\bibinfo {author} {\bibfnamefont {J.}~\bibnamefont
  {Nettersheim}}, \bibinfo {author} {\bibfnamefont {S.}~\bibnamefont
  {Burgardt}}, \bibinfo {author} {\bibfnamefont {Q.}~\bibnamefont {Bouton}},
  \bibinfo {author} {\bibfnamefont {D.}~\bibnamefont {Adam}}, \bibinfo {author}
  {\bibfnamefont {E.}~\bibnamefont {Lutz}},\ and\ \bibinfo {author}
  {\bibfnamefont {A.}~\bibnamefont {Widera}},\ }\bibfield  {title} {\bibinfo
  {title} {Power of a quasispin quantum otto engine at negative effective spin
  temperature},\ }\href {https://doi.org/10.1103/PRXQuantum.3.040334}
  {\bibfield  {journal} {\bibinfo  {journal} {PRX Quantum}\ }\textbf {\bibinfo
  {volume} {3}},\ \bibinfo {pages} {040334} (\bibinfo {year}
  {2022})}\BibitemShut {NoStop}%
\bibitem [{\citenamefont {Abah}\ \emph {et~al.}(2012)\citenamefont {Abah},
  \citenamefont {Ro\ss{}nagel}, \citenamefont {Jacob}, \citenamefont {Deffner},
  \citenamefont {Schmidt-Kaler}, \citenamefont {Singer},\ and\ \citenamefont
  {Lutz}}]{abah2012}%
  \BibitemOpen
  \bibfield  {author} {\bibinfo {author} {\bibfnamefont {O.}~\bibnamefont
  {Abah}}, \bibinfo {author} {\bibfnamefont {J.}~\bibnamefont {Ro\ss{}nagel}},
  \bibinfo {author} {\bibfnamefont {G.}~\bibnamefont {Jacob}}, \bibinfo
  {author} {\bibfnamefont {S.}~\bibnamefont {Deffner}}, \bibinfo {author}
  {\bibfnamefont {F.}~\bibnamefont {Schmidt-Kaler}}, \bibinfo {author}
  {\bibfnamefont {K.}~\bibnamefont {Singer}},\ and\ \bibinfo {author}
  {\bibfnamefont {E.}~\bibnamefont {Lutz}},\ }\bibfield  {title} {\bibinfo
  {title} {Single-ion heat engine at maximum power},\ }\href
  {https://doi.org/10.1103/PhysRevLett.109.203006} {\bibfield  {journal}
  {\bibinfo  {journal} {Phys. Rev. Lett.}\ }\textbf {\bibinfo {volume} {109}},\
  \bibinfo {pages} {203006} (\bibinfo {year} {2012})}\BibitemShut {NoStop}%
\bibitem [{\citenamefont {H{\"u}bner}\ \emph {et~al.}()\citenamefont
  {H{\"u}bner}, \citenamefont {Lefkidis}, \citenamefont {Dong}, \citenamefont
  {Chaudhuri}, \citenamefont {Chotorlishvili},\ and\ \citenamefont
  {Berakdar}}]{hubner2014}%
  \BibitemOpen
  \bibfield  {author} {\bibinfo {author} {\bibfnamefont {W.}~\bibnamefont
  {H{\"u}bner}}, \bibinfo {author} {\bibfnamefont {G.}~\bibnamefont
  {Lefkidis}}, \bibinfo {author} {\bibfnamefont {C.}~\bibnamefont {Dong}},
  \bibinfo {author} {\bibfnamefont {D.}~\bibnamefont {Chaudhuri}}, \bibinfo
  {author} {\bibfnamefont {L.}~\bibnamefont {Chotorlishvili}},\ and\ \bibinfo
  {author} {\bibfnamefont {J.}~\bibnamefont {Berakdar}},\ }\bibfield  {title}
  {\bibinfo {title} {Spin-dependent {O}tto quantum heat engine based on a
  molecular substance},\ }\href {https://doi.org/10.1103/PhysRevB.90.024401}
  {\bibfield  {journal} {\bibinfo  {journal} {Phys. Rev. B}\ }\textbf {\bibinfo
  {volume} {90}},\ \bibinfo {pages} {024401}}\BibitemShut {NoStop}%
\bibitem [{\citenamefont {Solfanelli}\ \emph {et~al.}(2020)\citenamefont
  {Solfanelli}, \citenamefont {Falsetti},\ and\ \citenamefont
  {Campisi}}]{PhysRevB.101.054513}%
  \BibitemOpen
  \bibfield  {author} {\bibinfo {author} {\bibfnamefont {A.}~\bibnamefont
  {Solfanelli}}, \bibinfo {author} {\bibfnamefont {M.}~\bibnamefont
  {Falsetti}},\ and\ \bibinfo {author} {\bibfnamefont {M.}~\bibnamefont
  {Campisi}},\ }\bibfield  {title} {\bibinfo {title} {Nonadiabatic single-qubit
  quantum {O}tto engine},\ }\href {https://doi.org/10.1103/PhysRevB.101.054513}
  {\bibfield  {journal} {\bibinfo  {journal} {Phys. Rev. B}\ }\textbf {\bibinfo
  {volume} {101}},\ \bibinfo {pages} {054513} (\bibinfo {year}
  {2020})}\BibitemShut {NoStop}%
\bibitem [{\citenamefont {Karimi}\ and\ \citenamefont
  {Pekola}(2016)}]{PhysRevB.94.184503}%
  \BibitemOpen
  \bibfield  {author} {\bibinfo {author} {\bibfnamefont {B.}~\bibnamefont
  {Karimi}}\ and\ \bibinfo {author} {\bibfnamefont {J.~P.}\ \bibnamefont
  {Pekola}},\ }\bibfield  {title} {\bibinfo {title} {Otto refrigerator based on
  a superconducting qubit: Classical and quantum performance},\ }\href
  {https://doi.org/10.1103/PhysRevB.94.184503} {\bibfield  {journal} {\bibinfo
  {journal} {Phys. Rev. B}\ }\textbf {\bibinfo {volume} {94}},\ \bibinfo
  {pages} {184503} (\bibinfo {year} {2016})}\BibitemShut {NoStop}%
\bibitem [{\citenamefont {Thomas}\ and\ \citenamefont
  {Pekola}(2023)}]{PhysRevResearch.5.L022036}%
  \BibitemOpen
  \bibfield  {author} {\bibinfo {author} {\bibfnamefont {G.}~\bibnamefont
  {Thomas}}\ and\ \bibinfo {author} {\bibfnamefont {J.~P.}\ \bibnamefont
  {Pekola}},\ }\bibfield  {title} {\bibinfo {title} {Dynamical phase and
  quantum heat at fractional frequencies},\ }\href
  {https://doi.org/10.1103/PhysRevResearch.5.L022036} {\bibfield  {journal}
  {\bibinfo  {journal} {Phys. Rev. Res.}\ }\textbf {\bibinfo {volume} {5}},\
  \bibinfo {pages} {L022036} (\bibinfo {year} {2023})}\BibitemShut {NoStop}%
\bibitem [{\citenamefont {Ji}\ \emph {et~al.}(2022)\citenamefont {Ji},
  \citenamefont {Chai}, \citenamefont {Wang}, \citenamefont {Guo},
  \citenamefont {Rong}, \citenamefont {Shi}, \citenamefont {Ren}, \citenamefont
  {Wang},\ and\ \citenamefont {Du}}]{PhysRevLett.128.090602}%
  \BibitemOpen
  \bibfield  {author} {\bibinfo {author} {\bibfnamefont {W.}~\bibnamefont
  {Ji}}, \bibinfo {author} {\bibfnamefont {Z.}~\bibnamefont {Chai}}, \bibinfo
  {author} {\bibfnamefont {M.}~\bibnamefont {Wang}}, \bibinfo {author}
  {\bibfnamefont {Y.}~\bibnamefont {Guo}}, \bibinfo {author} {\bibfnamefont
  {X.}~\bibnamefont {Rong}}, \bibinfo {author} {\bibfnamefont {F.}~\bibnamefont
  {Shi}}, \bibinfo {author} {\bibfnamefont {C.}~\bibnamefont {Ren}}, \bibinfo
  {author} {\bibfnamefont {Y.}~\bibnamefont {Wang}},\ and\ \bibinfo {author}
  {\bibfnamefont {J.}~\bibnamefont {Du}},\ }\bibfield  {title} {\bibinfo
  {title} {Spin quantum heat engine quantified by quantum steering},\ }\href
  {https://doi.org/10.1103/PhysRevLett.128.090602} {\bibfield  {journal}
  {\bibinfo  {journal} {Phys. Rev. Lett.}\ }\textbf {\bibinfo {volume} {128}},\
  \bibinfo {pages} {090602} (\bibinfo {year} {2022})}\BibitemShut {NoStop}%
\bibitem [{\citenamefont {Del~Grosso}\ \emph {et~al.}(2022)\citenamefont
  {Del~Grosso}, \citenamefont {Lombardo}, \citenamefont {Mazzitelli},\ and\
  \citenamefont {Villar}}]{QOCinasupercoducting}%
  \BibitemOpen
  \bibfield  {author} {\bibinfo {author} {\bibfnamefont {N.~F.}\ \bibnamefont
  {Del~Grosso}}, \bibinfo {author} {\bibfnamefont {F.~C.}\ \bibnamefont
  {Lombardo}}, \bibinfo {author} {\bibfnamefont {F.~D.}\ \bibnamefont
  {Mazzitelli}},\ and\ \bibinfo {author} {\bibfnamefont {P.~I.}\ \bibnamefont
  {Villar}},\ }\bibfield  {title} {\bibinfo {title} {Quantum {O}tto cycle in a
  superconducting cavity in the nonadiabatic regime},\ }\href
  {https://doi.org/10.1103/PhysRevA.105.022202} {\bibfield  {journal} {\bibinfo
   {journal} {Phys. Rev. A}\ }\textbf {\bibinfo {volume} {105}},\ \bibinfo
  {pages} {022202} (\bibinfo {year} {2022})}\BibitemShut {NoStop}%
\bibitem [{\citenamefont {Myers}\ and\ \citenamefont
  {Deffner}(2021)}]{Statistsanyone}%
  \BibitemOpen
  \bibfield  {author} {\bibinfo {author} {\bibfnamefont {N.~M.}\ \bibnamefont
  {Myers}}\ and\ \bibinfo {author} {\bibfnamefont {S.}~\bibnamefont
  {Deffner}},\ }\bibfield  {title} {\bibinfo {title} {Thermodynamics of
  statistical anyons},\ }\href {https://doi.org/10.1103/PRXQuantum.2.040312}
  {\bibfield  {journal} {\bibinfo  {journal} {PRX Quantum}\ }\textbf {\bibinfo
  {volume} {2}},\ \bibinfo {pages} {040312} (\bibinfo {year}
  {2021})}\BibitemShut {NoStop}%
\bibitem [{\citenamefont {Ono}\ \emph {et~al.}(2020)\citenamefont {Ono},
  \citenamefont {Shevchenko}, \citenamefont {Mori}, \citenamefont {Moriyama},\
  and\ \citenamefont {Nori}}]{PhysRevLett.125.166802}%
  \BibitemOpen
  \bibfield  {author} {\bibinfo {author} {\bibfnamefont {K.}~\bibnamefont
  {Ono}}, \bibinfo {author} {\bibfnamefont {S.~N.}\ \bibnamefont {Shevchenko}},
  \bibinfo {author} {\bibfnamefont {T.}~\bibnamefont {Mori}}, \bibinfo {author}
  {\bibfnamefont {S.}~\bibnamefont {Moriyama}},\ and\ \bibinfo {author}
  {\bibfnamefont {F.}~\bibnamefont {Nori}},\ }\bibfield  {title} {\bibinfo
  {title} {Analog of a quantum heat engine using a single-spin qubit},\ }\href
  {https://doi.org/10.1103/PhysRevLett.125.166802} {\bibfield  {journal}
  {\bibinfo  {journal} {Phys. Rev. Lett.}\ }\textbf {\bibinfo {volume} {125}},\
  \bibinfo {pages} {166802} (\bibinfo {year} {2020})}\BibitemShut {NoStop}%
\bibitem [{\citenamefont {Klatzow}\ \emph {et~al.}(2019)\citenamefont
  {Klatzow}, \citenamefont {Becker}, \citenamefont {Ledingham}, \citenamefont
  {Weinzetl}, \citenamefont {Kaczmarek}, \citenamefont {Saunders},
  \citenamefont {Nunn}, \citenamefont {Walmsley}, \citenamefont {Uzdin},\ and\
  \citenamefont {Poem}}]{PhysRevLett.122.110601}%
  \BibitemOpen
  \bibfield  {author} {\bibinfo {author} {\bibfnamefont {J.}~\bibnamefont
  {Klatzow}}, \bibinfo {author} {\bibfnamefont {J.~N.}\ \bibnamefont {Becker}},
  \bibinfo {author} {\bibfnamefont {P.~M.}\ \bibnamefont {Ledingham}}, \bibinfo
  {author} {\bibfnamefont {C.}~\bibnamefont {Weinzetl}}, \bibinfo {author}
  {\bibfnamefont {K.~T.}\ \bibnamefont {Kaczmarek}}, \bibinfo {author}
  {\bibfnamefont {D.~J.}\ \bibnamefont {Saunders}}, \bibinfo {author}
  {\bibfnamefont {J.}~\bibnamefont {Nunn}}, \bibinfo {author} {\bibfnamefont
  {I.~A.}\ \bibnamefont {Walmsley}}, \bibinfo {author} {\bibfnamefont
  {R.}~\bibnamefont {Uzdin}},\ and\ \bibinfo {author} {\bibfnamefont
  {E.}~\bibnamefont {Poem}},\ }\bibfield  {title} {\bibinfo {title}
  {Experimental demonstration of quantum effects in the operation of
  microscopic heat engines},\ }\href
  {https://doi.org/10.1103/PhysRevLett.122.110601} {\bibfield  {journal}
  {\bibinfo  {journal} {Phys. Rev. Lett.}\ }\textbf {\bibinfo {volume} {122}},\
  \bibinfo {pages} {110601} (\bibinfo {year} {2019})}\BibitemShut {NoStop}%
\bibitem [{\citenamefont {Zou}\ \emph {et~al.}(2017)\citenamefont {Zou},
  \citenamefont {Jiang}, \citenamefont {Mei}, \citenamefont {Guo},\ and\
  \citenamefont {Du}}]{EITQHE_2017}%
  \BibitemOpen
  \bibfield  {author} {\bibinfo {author} {\bibfnamefont {Y.}~\bibnamefont
  {Zou}}, \bibinfo {author} {\bibfnamefont {Y.}~\bibnamefont {Jiang}}, \bibinfo
  {author} {\bibfnamefont {Y.}~\bibnamefont {Mei}}, \bibinfo {author}
  {\bibfnamefont {X.}~\bibnamefont {Guo}},\ and\ \bibinfo {author}
  {\bibfnamefont {S.}~\bibnamefont {Du}},\ }\bibfield  {title} {\bibinfo
  {title} {Quantum heat engine using electromagnetically induced
  transparency},\ }\href {https://doi.org/10.1103/PhysRevLett.119.050602}
  {\bibfield  {journal} {\bibinfo  {journal} {Phys. Rev. Lett.}\ }\textbf
  {\bibinfo {volume} {119}},\ \bibinfo {pages} {050602} (\bibinfo {year}
  {2017})}\BibitemShut {NoStop}%
\bibitem [{\citenamefont {von Lindenfels}\ \emph {et~al.}(2019)\citenamefont
  {von Lindenfels}, \citenamefont {Gr\"ab}, \citenamefont {Schmiegelow},
  \citenamefont {Kaushal}, \citenamefont {Schulz}, \citenamefont {Mitchison},
  \citenamefont {Goold}, \citenamefont {Schmidt-Kaler},\ and\ \citenamefont
  {Poschinger}}]{flywheel_2019}%
  \BibitemOpen
  \bibfield  {author} {\bibinfo {author} {\bibfnamefont {D.}~\bibnamefont {von
  Lindenfels}}, \bibinfo {author} {\bibfnamefont {O.}~\bibnamefont {Gr\"ab}},
  \bibinfo {author} {\bibfnamefont {C.~T.}\ \bibnamefont {Schmiegelow}},
  \bibinfo {author} {\bibfnamefont {V.}~\bibnamefont {Kaushal}}, \bibinfo
  {author} {\bibfnamefont {J.}~\bibnamefont {Schulz}}, \bibinfo {author}
  {\bibfnamefont {M.~T.}\ \bibnamefont {Mitchison}}, \bibinfo {author}
  {\bibfnamefont {J.}~\bibnamefont {Goold}}, \bibinfo {author} {\bibfnamefont
  {F.}~\bibnamefont {Schmidt-Kaler}},\ and\ \bibinfo {author} {\bibfnamefont
  {U.~G.}\ \bibnamefont {Poschinger}},\ }\bibfield  {title} {\bibinfo {title}
  {Spin heat engine coupled to a harmonic-oscillator flywheel},\ }\href
  {https://doi.org/10.1103/PhysRevLett.123.080602} {\bibfield  {journal}
  {\bibinfo  {journal} {Phys. Rev. Lett.}\ }\textbf {\bibinfo {volume} {123}},\
  \bibinfo {pages} {080602} (\bibinfo {year} {2019})}\BibitemShut {NoStop}%
\bibitem [{\citenamefont {Ro{\ss}nagel}\ \emph {et~al.}(2016)\citenamefont
  {Ro{\ss}nagel}, \citenamefont {Dawkins}, \citenamefont {Tolazzi},
  \citenamefont {Abah}, \citenamefont {Lutz}, \citenamefont {Schmidt-Kaler},\
  and\ \citenamefont {Singer}}]{Ro_nagel_2016}%
  \BibitemOpen
  \bibfield  {author} {\bibinfo {author} {\bibfnamefont {J.}~\bibnamefont
  {Ro{\ss}nagel}}, \bibinfo {author} {\bibfnamefont {S.~T.}\ \bibnamefont
  {Dawkins}}, \bibinfo {author} {\bibfnamefont {K.~N.}\ \bibnamefont
  {Tolazzi}}, \bibinfo {author} {\bibfnamefont {O.}~\bibnamefont {Abah}},
  \bibinfo {author} {\bibfnamefont {E.}~\bibnamefont {Lutz}}, \bibinfo {author}
  {\bibfnamefont {F.}~\bibnamefont {Schmidt-Kaler}},\ and\ \bibinfo {author}
  {\bibfnamefont {K.}~\bibnamefont {Singer}},\ }\bibfield  {title} {\bibinfo
  {title} {A single-atom heat engine},\ }\href
  {https://doi.org/10.1126/science.aad6320} {\bibfield  {journal} {\bibinfo
  {journal} {Science}\ }\textbf {\bibinfo {volume} {352}},\ \bibinfo {pages}
  {325} (\bibinfo {year} {2016})}\BibitemShut {NoStop}%
\bibitem [{\citenamefont {Maslennikov}\ \emph {et~al.}(2019)\citenamefont
  {Maslennikov}, \citenamefont {Ding}, \citenamefont {Hablützel},
  \citenamefont {Gan}, \citenamefont {Roulet}, \citenamefont {Nimmrichter},
  \citenamefont {Dai}, \citenamefont {Scarani},\ and\ \citenamefont
  {Matsukevich}}]{Maslennikov_2019}%
  \BibitemOpen
  \bibfield  {author} {\bibinfo {author} {\bibfnamefont {G.}~\bibnamefont
  {Maslennikov}}, \bibinfo {author} {\bibfnamefont {S.}~\bibnamefont {Ding}},
  \bibinfo {author} {\bibfnamefont {R.}~\bibnamefont {Hablützel}}, \bibinfo
  {author} {\bibfnamefont {J.}~\bibnamefont {Gan}}, \bibinfo {author}
  {\bibfnamefont {A.}~\bibnamefont {Roulet}}, \bibinfo {author} {\bibfnamefont
  {S.}~\bibnamefont {Nimmrichter}}, \bibinfo {author} {\bibfnamefont
  {J.}~\bibnamefont {Dai}}, \bibinfo {author} {\bibfnamefont {V.}~\bibnamefont
  {Scarani}},\ and\ \bibinfo {author} {\bibfnamefont {D.}~\bibnamefont
  {Matsukevich}},\ }\bibfield  {title} {\bibinfo {title} {Quantum absorption
  refrigerator with trapped ions},\ }\href
  {https://doi.org/10.1038/s41467-018-08090-0} {\bibfield  {journal} {\bibinfo
  {journal} {Nature Communications}\ }\textbf {\bibinfo {volume} {10}},\
  \bibinfo {pages} {202} (\bibinfo {year} {2019})}\BibitemShut {NoStop}%
\bibitem [{\citenamefont {Bouton}\ \emph {et~al.}(2021)\citenamefont {Bouton},
  \citenamefont {Nettersheim}, \citenamefont {Burgardt}, \citenamefont {Adam},
  \citenamefont {Lutz},\ and\ \citenamefont {Widera}}]{Bouton_2021}%
  \BibitemOpen
  \bibfield  {author} {\bibinfo {author} {\bibfnamefont {Q.}~\bibnamefont
  {Bouton}}, \bibinfo {author} {\bibfnamefont {J.}~\bibnamefont {Nettersheim}},
  \bibinfo {author} {\bibfnamefont {S.}~\bibnamefont {Burgardt}}, \bibinfo
  {author} {\bibfnamefont {D.}~\bibnamefont {Adam}}, \bibinfo {author}
  {\bibfnamefont {E.}~\bibnamefont {Lutz}},\ and\ \bibinfo {author}
  {\bibfnamefont {A.}~\bibnamefont {Widera}},\ }\bibfield  {title} {\bibinfo
  {title} {A quantum heat engine driven by atomic collisions},\ }\href
  {https://doi.org/10.1038/s41467-021-22222-z} {\bibfield  {journal} {\bibinfo
  {journal} {Nature Communications}\ }\textbf {\bibinfo {volume} {12}},\
  \bibinfo {pages} {2063} (\bibinfo {year} {2021})}\BibitemShut {NoStop}%
\bibitem [{\citenamefont {Beyer}\ \emph {et~al.}(2019)\citenamefont {Beyer},
  \citenamefont {Luoma},\ and\ \citenamefont
  {Strunz}}]{PhysRevLett.123.250606}%
  \BibitemOpen
  \bibfield  {author} {\bibinfo {author} {\bibfnamefont {K.}~\bibnamefont
  {Beyer}}, \bibinfo {author} {\bibfnamefont {K.}~\bibnamefont {Luoma}},\ and\
  \bibinfo {author} {\bibfnamefont {W.~T.}\ \bibnamefont {Strunz}},\ }\bibfield
   {title} {\bibinfo {title} {Steering heat engines: A truly quantum {M}axwell
  demon},\ }\href {https://doi.org/10.1103/PhysRevLett.123.250606} {\bibfield
  {journal} {\bibinfo  {journal} {Phys. Rev. Lett.}\ }\textbf {\bibinfo
  {volume} {123}},\ \bibinfo {pages} {250606} (\bibinfo {year}
  {2019})}\BibitemShut {NoStop}%
\bibitem [{\citenamefont {Altintas}\ and\ \citenamefont {M\"ustecapl\ifmmode
  \imath \else \i \fi{}o\ifmmode~\breve{g}\else
  \u{g}\fi{}lu}(2015)}]{Altintas2015}%
  \BibitemOpen
  \bibfield  {author} {\bibinfo {author} {\bibfnamefont {F.}~\bibnamefont
  {Altintas}}\ and\ \bibinfo {author} {\bibfnamefont {O.~E.}\ \bibnamefont
  {M\"ustecapl\ifmmode \imath \else \i \fi{}o\ifmmode~\breve{g}\else
  \u{g}\fi{}lu}},\ }\bibfield  {title} {\bibinfo {title} {General formalism of
  local thermodynamics with an example: Quantum {O}tto engine with a spin-$1/2$
  coupled to an arbitrary spin},\ }\href
  {https://doi.org/10.1103/PhysRevE.92.022142} {\bibfield  {journal} {\bibinfo
  {journal} {Phys. Rev. E}\ }\textbf {\bibinfo {volume} {92}},\ \bibinfo
  {pages} {022142} (\bibinfo {year} {2015})}\BibitemShut {NoStop}%
\bibitem [{\citenamefont {Thomas}\ and\ \citenamefont
  {Johal}(2011)}]{Thomas2011}%
  \BibitemOpen
  \bibfield  {author} {\bibinfo {author} {\bibfnamefont {G.}~\bibnamefont
  {Thomas}}\ and\ \bibinfo {author} {\bibfnamefont {R.~S.}\ \bibnamefont
  {Johal}},\ }\bibfield  {title} {\bibinfo {title} {Coupled quantum {O}tto
  cycle},\ }\href {https://doi.org/10.1103/PhysRevE.83.031135} {\bibfield
  {journal} {\bibinfo  {journal} {Phys. Rev. E}\ }\textbf {\bibinfo {volume}
  {83}},\ \bibinfo {pages} {031135} (\bibinfo {year} {2011})}\BibitemShut
  {NoStop}%
\bibitem [{\citenamefont {Thomas}\ and\ \citenamefont
  {Johal}(2014)}]{2014friction}%
  \BibitemOpen
  \bibfield  {author} {\bibinfo {author} {\bibfnamefont {G.}~\bibnamefont
  {Thomas}}\ and\ \bibinfo {author} {\bibfnamefont {R.~S.}\ \bibnamefont
  {Johal}},\ }\bibfield  {title} {\bibinfo {title} {Friction due to
  inhomogeneous driving of coupled spins in a quantum heat engine},\ }\href
  {https://doi.org/https://doi.org/10.1140/epjb/e2014-50231-1} {\bibfield
  {journal} {\bibinfo  {journal} {The European Phys. Journal B}\ }\textbf
  {\bibinfo {volume} {87}},\ \bibinfo {pages} {166} (\bibinfo {year}
  {2014})}\BibitemShut {NoStop}%
\bibitem [{\citenamefont {Johal}\ and\ \citenamefont {Mehta}(2021)}]{venu2021}%
  \BibitemOpen
  \bibfield  {author} {\bibinfo {author} {\bibfnamefont {R.~S.}\ \bibnamefont
  {Johal}}\ and\ \bibinfo {author} {\bibfnamefont {V.}~\bibnamefont {Mehta}},\
  }\bibfield  {title} {\bibinfo {title} {Quantum heat engines with complex
  working media, complete {O}tto cycles and heuristics},\ }\href
  {https://doi.org/10.3390/e23091149} {\bibfield  {journal} {\bibinfo
  {journal} {Entropy}\ }\textbf {\bibinfo {volume} {23}},\ \bibinfo {pages}
  {1149} (\bibinfo {year} {2021})}\BibitemShut {NoStop}%
\bibitem [{\citenamefont {Sonkar}\ and\ \citenamefont
  {Johal}(2023)}]{Sachin2023}%
  \BibitemOpen
  \bibfield  {author} {\bibinfo {author} {\bibfnamefont {S.}~\bibnamefont
  {Sonkar}}\ and\ \bibinfo {author} {\bibfnamefont {R.~S.}\ \bibnamefont
  {Johal}},\ }\bibfield  {title} {\bibinfo {title} {Spin-based quantum {O}tto
  engines and majorization},\ }\href
  {https://doi.org/10.1103/PhysRevA.107.032220} {\bibfield  {journal} {\bibinfo
   {journal} {Phys. Rev. A}\ }\textbf {\bibinfo {volume} {107}},\ \bibinfo
  {pages} {032220} (\bibinfo {year} {2023})}\BibitemShut {NoStop}%
\bibitem [{\citenamefont {Ivanchenko}(2015)}]{Ivanchenko2015}%
  \BibitemOpen
  \bibfield  {author} {\bibinfo {author} {\bibfnamefont {E.~A.}\ \bibnamefont
  {Ivanchenko}},\ }\bibfield  {title} {\bibinfo {title} {Quantum {O}tto cycle
  efficiency on coupled qudits},\ }\href
  {https://doi.org/10.1103/PhysRevE.92.032124} {\bibfield  {journal} {\bibinfo
  {journal} {Phys. Rev. E}\ }\textbf {\bibinfo {volume} {92}},\ \bibinfo
  {pages} {032124} (\bibinfo {year} {2015})}\BibitemShut {NoStop}%
\bibitem [{\citenamefont {T{\"u}rkpen{\c{c}}e}\ and\ \citenamefont
  {Altintas}(2019)}]{turkpencce2019coupled}%
  \BibitemOpen
  \bibfield  {author} {\bibinfo {author} {\bibfnamefont {D.}~\bibnamefont
  {T{\"u}rkpen{\c{c}}e}}\ and\ \bibinfo {author} {\bibfnamefont
  {F.}~\bibnamefont {Altintas}},\ }\bibfield  {title} {\bibinfo {title}
  {Coupled quantum {O}tto heat engine and refrigerator with inner friction},\
  }\href {https://doi.org/10.1007/s11128-019-2366-7} {\bibfield  {journal}
  {\bibinfo  {journal} {Quantum Information Processing}\ }\textbf {\bibinfo
  {volume} {18}},\ \bibinfo {pages} {255} (\bibinfo {year} {2019})}\BibitemShut
  {NoStop}%
\bibitem [{\citenamefont {Brunner}\ \emph {et~al.}(2014)\citenamefont
  {Brunner}, \citenamefont {Huber}, \citenamefont {Linden}, \citenamefont
  {Popescu}, \citenamefont {Silva},\ and\ \citenamefont
  {Skrzypczyk}}]{Brunner2014}%
  \BibitemOpen
  \bibfield  {author} {\bibinfo {author} {\bibfnamefont {N.}~\bibnamefont
  {Brunner}}, \bibinfo {author} {\bibfnamefont {M.}~\bibnamefont {Huber}},
  \bibinfo {author} {\bibfnamefont {N.}~\bibnamefont {Linden}}, \bibinfo
  {author} {\bibfnamefont {S.}~\bibnamefont {Popescu}}, \bibinfo {author}
  {\bibfnamefont {R.}~\bibnamefont {Silva}},\ and\ \bibinfo {author}
  {\bibfnamefont {P.}~\bibnamefont {Skrzypczyk}},\ }\bibfield  {title}
  {\bibinfo {title} {Entanglement enhances cooling in microscopic quantum
  refrigerators},\ }\href {https://doi.org/10.1103/PhysRevE.89.032115}
  {\bibfield  {journal} {\bibinfo  {journal} {Phys. Rev. E}\ }\textbf {\bibinfo
  {volume} {89}},\ \bibinfo {pages} {032115} (\bibinfo {year}
  {2014})}\BibitemShut {NoStop}%
\bibitem [{\citenamefont {Zhang}(2020)}]{zhang2019optimization}%
  \BibitemOpen
  \bibfield  {author} {\bibinfo {author} {\bibfnamefont {Y.}~\bibnamefont
  {Zhang}},\ }\bibfield  {title} {\bibinfo {title} {Optimization performance of
  quantum {O}tto heat engines and refrigerators with squeezed thermal
  reservoirs},\ }\href
  {https://doi.org/https://doi.org/10.1016/j.physa.2020.125083} {\bibfield
  {journal} {\bibinfo  {journal} {Physica A: Statistical Mechanics and its
  Applications}\ }\textbf {\bibinfo {volume} {559}},\ \bibinfo {pages} {125083}
  (\bibinfo {year} {2020})}\BibitemShut {NoStop}%
\bibitem [{\citenamefont {de~Assis}\ \emph {et~al.}(2021)\citenamefont
  {de~Assis}, \citenamefont {Sales}, \citenamefont {Mendes},\ and\
  \citenamefont {de~Almeida}}]{de_Assis_2021}%
  \BibitemOpen
  \bibfield  {author} {\bibinfo {author} {\bibfnamefont {R.~J.}\ \bibnamefont
  {de~Assis}}, \bibinfo {author} {\bibfnamefont {J.~S.}\ \bibnamefont {Sales}},
  \bibinfo {author} {\bibfnamefont {U.~C.}\ \bibnamefont {Mendes}},\ and\
  \bibinfo {author} {\bibfnamefont {N.~G.}\ \bibnamefont {de~Almeida}},\
  }\bibfield  {title} {\bibinfo {title} {Two-level quantum {O}tto heat engine
  operating with unit efficiency far from the quasi-static regime under a
  squeezed reservoir},\ }\href {https://doi.org/10.1088/1361-6455/abcfd9}
  {\bibfield  {journal} {\bibinfo  {journal} {Journal of Physics B: Atomic,
  Molecular and Optical Physics}\ }\textbf {\bibinfo {volume} {54}},\ \bibinfo
  {pages} {095501} (\bibinfo {year} {2021})}\BibitemShut {NoStop}%
\bibitem [{\citenamefont {Singh}\ and\ \citenamefont {M\"ustecapl\ifmmode
  \imath \else \i \fi{}o\ifmmode~\breve{g}\else
  \u{g}\fi{}lu}(2020)}]{Singh_2020}%
  \BibitemOpen
  \bibfield  {author} {\bibinfo {author} {\bibfnamefont {V.}~\bibnamefont
  {Singh}}\ and\ \bibinfo {author} {\bibfnamefont {O.~E.}\ \bibnamefont
  {M\"ustecapl\ifmmode \imath \else \i \fi{}o\ifmmode~\breve{g}\else
  \u{g}\fi{}lu}},\ }\bibfield  {title} {\bibinfo {title} {Performance bounds of
  nonadiabatic quantum harmonic {O}tto engine and refrigerator under a squeezed
  thermal reservoir},\ }\href {https://doi.org/10.1103/physreve.102.062123}
  {\bibfield  {journal} {\bibinfo  {journal} {Physical Review E}\ }\textbf
  {\bibinfo {volume} {102}},\ \bibinfo {pages} {062123} (\bibinfo {year}
  {2020})}\BibitemShut {NoStop}%
\bibitem [{\citenamefont {Wang}\ \emph {et~al.}(2019)\citenamefont {Wang},
  \citenamefont {He},\ and\ \citenamefont {Ma}}]{wwang2019}%
  \BibitemOpen
  \bibfield  {author} {\bibinfo {author} {\bibfnamefont {J.}~\bibnamefont
  {Wang}}, \bibinfo {author} {\bibfnamefont {J.}~\bibnamefont {He}},\ and\
  \bibinfo {author} {\bibfnamefont {Y.}~\bibnamefont {Ma}},\ }\bibfield
  {title} {\bibinfo {title} {Finite-time performance of a quantum heat engine
  with a squeezed thermal bath},\ }\href
  {https://doi.org/10.1103/PhysRevE.100.052126} {\bibfield  {journal} {\bibinfo
   {journal} {Phys. Rev. E}\ }\textbf {\bibinfo {volume} {100}},\ \bibinfo
  {pages} {052126} (\bibinfo {year} {2019})}\BibitemShut {NoStop}%
\bibitem [{\citenamefont {Huang}\ \emph {et~al.}(2017)\citenamefont {Huang},
  \citenamefont {Guo}, \citenamefont {Wu},\ and\ \citenamefont
  {Yi}}]{Huang_2017}%
  \BibitemOpen
  \bibfield  {author} {\bibinfo {author} {\bibfnamefont {X.~L.}\ \bibnamefont
  {Huang}}, \bibinfo {author} {\bibfnamefont {D.~Y.}\ \bibnamefont {Guo}},
  \bibinfo {author} {\bibfnamefont {S.~L.}\ \bibnamefont {Wu}},\ and\ \bibinfo
  {author} {\bibfnamefont {X.~X.}\ \bibnamefont {Yi}},\ }\bibfield  {title}
  {\bibinfo {title} {Multilevel quantum {O}tto heat engines with identical
  particles},\ }\href {https://doi.org/10.1007/s11128-017-1795-4} {\bibfield
  {journal} {\bibinfo  {journal} {Quantum Information Processing}\ }\textbf
  {\bibinfo {volume} {17}},\ \bibinfo {pages} {27} (\bibinfo {year}
  {2017})}\BibitemShut {NoStop}%
\bibitem [{\citenamefont {Wang}\ \emph {et~al.}(2012)\citenamefont {Wang},
  \citenamefont {Wu},\ and\ \citenamefont {He}}]{wong2012}%
  \BibitemOpen
  \bibfield  {author} {\bibinfo {author} {\bibfnamefont {J.}~\bibnamefont
  {Wang}}, \bibinfo {author} {\bibfnamefont {Z.}~\bibnamefont {Wu}},\ and\
  \bibinfo {author} {\bibfnamefont {J.}~\bibnamefont {He}},\ }\bibfield
  {title} {\bibinfo {title} {Quantum {O}tto engine of a two-level atom with
  single-mode fields},\ }\href {https://doi.org/10.1103/PhysRevE.85.041148}
  {\bibfield  {journal} {\bibinfo  {journal} {Phys. Rev. E}\ }\textbf {\bibinfo
  {volume} {85}},\ \bibinfo {pages} {041148} (\bibinfo {year}
  {2012})}\BibitemShut {NoStop}%
\bibitem [{\citenamefont {Yunger~Halpern}\ \emph {et~al.}(2019)\citenamefont
  {Yunger~Halpern}, \citenamefont {White}, \citenamefont {Gopalakrishnan},\
  and\ \citenamefont {Refael}}]{PhysRevB.99.024203}%
  \BibitemOpen
  \bibfield  {author} {\bibinfo {author} {\bibfnamefont {N.}~\bibnamefont
  {Yunger~Halpern}}, \bibinfo {author} {\bibfnamefont {C.~D.}\ \bibnamefont
  {White}}, \bibinfo {author} {\bibfnamefont {S.}~\bibnamefont
  {Gopalakrishnan}},\ and\ \bibinfo {author} {\bibfnamefont {G.}~\bibnamefont
  {Refael}},\ }\bibfield  {title} {\bibinfo {title} {Quantum engine based on
  many-body localization},\ }\href {https://doi.org/10.1103/PhysRevB.99.024203}
  {\bibfield  {journal} {\bibinfo  {journal} {Phys. Rev. B}\ }\textbf {\bibinfo
  {volume} {99}},\ \bibinfo {pages} {024203} (\bibinfo {year}
  {2019})}\BibitemShut {NoStop}%
\bibitem [{\citenamefont {Izadyari}\ \emph {et~al.}(2023)\citenamefont
  {Izadyari}, \citenamefont {Naseem},\ and\ \citenamefont {M\"ustecapl\ifmmode
  \imath \else \i \fi{}o\ifmmode~\breve{g}\else \u{g}\fi{}lu}}]{osgur_2023}%
  \BibitemOpen
  \bibfield  {author} {\bibinfo {author} {\bibfnamefont {M.}~\bibnamefont
  {Izadyari}}, \bibinfo {author} {\bibfnamefont {M.~T.}\ \bibnamefont
  {Naseem}},\ and\ \bibinfo {author} {\bibfnamefont {O.~E.}\ \bibnamefont
  {M\"ustecapl\ifmmode \imath \else \i \fi{}o\ifmmode~\breve{g}\else
  \u{g}\fi{}lu}},\ }\bibfield  {title} {\bibinfo {title} {Enantiomer detection
  via quantum {O}tto cycle},\ }\href
  {https://doi.org/10.1103/PhysRevE.107.L042103} {\bibfield  {journal}
  {\bibinfo  {journal} {Phys. Rev. E}\ }\textbf {\bibinfo {volume} {107}},\
  \bibinfo {pages} {L042103} (\bibinfo {year} {2023})}\BibitemShut {NoStop}%
\bibitem [{\citenamefont {Makouri}\ \emph {et~al.}(2023)\citenamefont
  {Makouri}, \citenamefont {Slaoui},\ and\ \citenamefont
  {Daoud}}]{El_Makouri_2023}%
  \BibitemOpen
  \bibfield  {author} {\bibinfo {author} {\bibfnamefont {A.~E.}\ \bibnamefont
  {Makouri}}, \bibinfo {author} {\bibfnamefont {A.}~\bibnamefont {Slaoui}},\
  and\ \bibinfo {author} {\bibfnamefont {M.}~\bibnamefont {Daoud}},\ }\bibfield
   {title} {\bibinfo {title} {Enhancing the performance of coupled quantum
  {O}tto thermal machines without entanglement and quantum correlations},\
  }\href {https://doi.org/10.1088/1361-6455/acc36d} {\bibfield  {journal}
  {\bibinfo  {journal} {Journal of Physics B: Atomic, Molecular and Optical
  Physics}\ }\textbf {\bibinfo {volume} {56}},\ \bibinfo {pages} {085501}
  (\bibinfo {year} {2023})}\BibitemShut {NoStop}%
\bibitem [{\citenamefont {Kaneyasu}\ and\ \citenamefont
  {Hasegawa}(2023)}]{hasegawa_2023}%
  \BibitemOpen
  \bibfield  {author} {\bibinfo {author} {\bibfnamefont {M.}~\bibnamefont
  {Kaneyasu}}\ and\ \bibinfo {author} {\bibfnamefont {Y.}~\bibnamefont
  {Hasegawa}},\ }\bibfield  {title} {\bibinfo {title} {Quantum {O}tto cycle
  under strong coupling},\ }\href {https://doi.org/10.1103/PhysRevE.107.044127}
  {\bibfield  {journal} {\bibinfo  {journal} {Phys. Rev. E}\ }\textbf {\bibinfo
  {volume} {107}},\ \bibinfo {pages} {044127} (\bibinfo {year}
  {2023})}\BibitemShut {NoStop}%
\bibitem [{\citenamefont {Piccitto}\ \emph {et~al.}(2022)\citenamefont
  {Piccitto}, \citenamefont {Campisi},\ and\ \citenamefont
  {Rossini}}]{Piccitto_2022}%
  \BibitemOpen
  \bibfield  {author} {\bibinfo {author} {\bibfnamefont {G.}~\bibnamefont
  {Piccitto}}, \bibinfo {author} {\bibfnamefont {M.}~\bibnamefont {Campisi}},\
  and\ \bibinfo {author} {\bibfnamefont {D.}~\bibnamefont {Rossini}},\
  }\bibfield  {title} {\bibinfo {title} {The {I}sing critical quantum {O}tto
  engine},\ }\href {https://doi.org/10.1088/1367-2630/ac963b} {\bibfield
  {journal} {\bibinfo  {journal} {New Journal of Physics}\ }\textbf {\bibinfo
  {volume} {24}},\ \bibinfo {pages} {103023} (\bibinfo {year}
  {2022})}\BibitemShut {NoStop}%
\bibitem [{\citenamefont {Solfanelli}\ \emph {et~al.}(2023)\citenamefont
  {Solfanelli}, \citenamefont {Giachetti}, \citenamefont {Campisi},
  \citenamefont {Ruffo},\ and\ \citenamefont {Defenu}}]{Solfanelli_2023}%
  \BibitemOpen
  \bibfield  {author} {\bibinfo {author} {\bibfnamefont {A.}~\bibnamefont
  {Solfanelli}}, \bibinfo {author} {\bibfnamefont {G.}~\bibnamefont
  {Giachetti}}, \bibinfo {author} {\bibfnamefont {M.}~\bibnamefont {Campisi}},
  \bibinfo {author} {\bibfnamefont {S.}~\bibnamefont {Ruffo}},\ and\ \bibinfo
  {author} {\bibfnamefont {N.}~\bibnamefont {Defenu}},\ }\bibfield  {title}
  {\bibinfo {title} {Quantum heat engine with long-range advantages},\ }\href
  {https://doi.org/10.1088/1367-2630/acc04e} {\bibfield  {journal} {\bibinfo
  {journal} {New Journal of Physics}\ }\textbf {\bibinfo {volume} {25}},\
  \bibinfo {pages} {033030} (\bibinfo {year} {2023})}\BibitemShut {NoStop}%
\bibitem [{\citenamefont {Ptaszy\ifmmode~\acute{n}\else
  \'{n}\fi{}ski}(2022)}]{PhysRevE.106.014114}%
  \BibitemOpen
  \bibfield  {author} {\bibinfo {author} {\bibfnamefont {K.}~\bibnamefont
  {Ptaszy\ifmmode~\acute{n}\else \'{n}\fi{}ski}},\ }\bibfield  {title}
  {\bibinfo {title} {Non-markovian thermal operations boosting the performance
  of quantum heat engines},\ }\href
  {https://doi.org/10.1103/PhysRevE.106.014114} {\bibfield  {journal} {\bibinfo
   {journal} {Phys. Rev. E}\ }\textbf {\bibinfo {volume} {106}},\ \bibinfo
  {pages} {014114} (\bibinfo {year} {2022})}\BibitemShut {NoStop}%
\bibitem [{\citenamefont {Ishizaki}\ \emph {et~al.}(2023)\citenamefont
  {Ishizaki}, \citenamefont {Hatano},\ and\ \citenamefont
  {Tajima}}]{PhysRevResearch.5.023066}%
  \BibitemOpen
  \bibfield  {author} {\bibinfo {author} {\bibfnamefont {M.}~\bibnamefont
  {Ishizaki}}, \bibinfo {author} {\bibfnamefont {N.}~\bibnamefont {Hatano}},\
  and\ \bibinfo {author} {\bibfnamefont {H.}~\bibnamefont {Tajima}},\
  }\bibfield  {title} {\bibinfo {title} {Switching the function of the quantum
  {O}tto cycle in non-markovian dynamics: Heat engine, heater, and heat pump},\
  }\href {https://doi.org/10.1103/PhysRevResearch.5.023066} {\bibfield
  {journal} {\bibinfo  {journal} {Phys. Rev. Res.}\ }\textbf {\bibinfo {volume}
  {5}},\ \bibinfo {pages} {023066} (\bibinfo {year} {2023})}\BibitemShut
  {NoStop}%
\bibitem [{\citenamefont {Chakraborty}\ \emph {et~al.}(2022)\citenamefont
  {Chakraborty}, \citenamefont {Das},\ and\ \citenamefont {Chru\ifmmode
  \acute{s}\else \'{s}\fi{}ci\ifmmode~\acute{n}\else
  \'{n}\fi{}ski}}]{sangik2022}%
  \BibitemOpen
  \bibfield  {author} {\bibinfo {author} {\bibfnamefont {S.}~\bibnamefont
  {Chakraborty}}, \bibinfo {author} {\bibfnamefont {A.}~\bibnamefont {Das}},\
  and\ \bibinfo {author} {\bibfnamefont {D.}~\bibnamefont {Chru\ifmmode
  \acute{s}\else \'{s}\fi{}ci\ifmmode~\acute{n}\else \'{n}\fi{}ski}},\
  }\bibfield  {title} {\bibinfo {title} {Strongly coupled quantum {O}tto cycle
  with single qubit bath},\ }\href
  {https://doi.org/10.1103/PhysRevE.106.064133} {\bibfield  {journal} {\bibinfo
   {journal} {Phys. Rev. E}\ }\textbf {\bibinfo {volume} {106}},\ \bibinfo
  {pages} {064133} (\bibinfo {year} {2022})}\BibitemShut {NoStop}%
\bibitem [{\citenamefont {Thomas}\ \emph {et~al.}(2018)\citenamefont {Thomas},
  \citenamefont {Siddharth}, \citenamefont {Banerjee},\ and\ \citenamefont
  {Ghosh}}]{Thomas2018}%
  \BibitemOpen
  \bibfield  {author} {\bibinfo {author} {\bibfnamefont {G.}~\bibnamefont
  {Thomas}}, \bibinfo {author} {\bibfnamefont {N.}~\bibnamefont {Siddharth}},
  \bibinfo {author} {\bibfnamefont {S.}~\bibnamefont {Banerjee}},\ and\
  \bibinfo {author} {\bibfnamefont {S.}~\bibnamefont {Ghosh}},\ }\bibfield
  {title} {\bibinfo {title} {Thermodynamics of non-markovian reservoirs and
  heat engines},\ }\href {https://doi.org/10.1103/PhysRevE.97.062108}
  {\bibfield  {journal} {\bibinfo  {journal} {Phys. Rev. E}\ }\textbf {\bibinfo
  {volume} {97}},\ \bibinfo {pages} {062108} (\bibinfo {year}
  {2018})}\BibitemShut {NoStop}%
\bibitem [{\citenamefont {Wu}\ \emph {et~al.}(2014)\citenamefont {Wu},
  \citenamefont {He}, \citenamefont {Ma},\ and\ \citenamefont {Wang}}]{wu2014}%
  \BibitemOpen
  \bibfield  {author} {\bibinfo {author} {\bibfnamefont {F.}~\bibnamefont
  {Wu}}, \bibinfo {author} {\bibfnamefont {J.}~\bibnamefont {He}}, \bibinfo
  {author} {\bibfnamefont {Y.}~\bibnamefont {Ma}},\ and\ \bibinfo {author}
  {\bibfnamefont {J.}~\bibnamefont {Wang}},\ }\bibfield  {title} {\bibinfo
  {title} {Efficiency at maximum power of a quantum {O}tto cycle within
  finite-time or irreversible thermodynamics},\ }\href
  {https://doi.org/10.1103/PhysRevE.90.062134} {\bibfield  {journal} {\bibinfo
  {journal} {Phys. Rev. E}\ }\textbf {\bibinfo {volume} {90}},\ \bibinfo
  {pages} {062134} (\bibinfo {year} {2014})}\BibitemShut {NoStop}%
\bibitem [{\citenamefont {Das}\ and\ \citenamefont
  {Mukherjee}(2020)}]{das2020}%
  \BibitemOpen
  \bibfield  {author} {\bibinfo {author} {\bibfnamefont {A.}~\bibnamefont
  {Das}}\ and\ \bibinfo {author} {\bibfnamefont {V.}~\bibnamefont
  {Mukherjee}},\ }\bibfield  {title} {\bibinfo {title} {Quantum-enhanced
  finite-time otto cycle},\ }\href
  {https://doi.org/10.1103/PhysRevResearch.2.033083} {\bibfield  {journal}
  {\bibinfo  {journal} {Phys. Rev. Res.}\ }\textbf {\bibinfo {volume} {2}},\
  \bibinfo {pages} {033083} (\bibinfo {year} {2020})}\BibitemShut {NoStop}%
\bibitem [{\citenamefont {Chand}\ \emph {et~al.}(2021)\citenamefont {Chand},
  \citenamefont {Dasgupta},\ and\ \citenamefont {Biswas}}]{chand2021}%
  \BibitemOpen
  \bibfield  {author} {\bibinfo {author} {\bibfnamefont {S.}~\bibnamefont
  {Chand}}, \bibinfo {author} {\bibfnamefont {S.}~\bibnamefont {Dasgupta}},\
  and\ \bibinfo {author} {\bibfnamefont {A.}~\bibnamefont {Biswas}},\
  }\bibfield  {title} {\bibinfo {title} {Finite-time performance of a
  single-ion quantum otto engine},\ }\href
  {https://doi.org/10.1103/PhysRevE.103.032144} {\bibfield  {journal} {\bibinfo
   {journal} {Phys. Rev. E}\ }\textbf {\bibinfo {volume} {103}},\ \bibinfo
  {pages} {032144} (\bibinfo {year} {2021})}\BibitemShut {NoStop}%
\bibitem [{\citenamefont {Lee}\ \emph {et~al.}(2020)\citenamefont {Lee},
  \citenamefont {Ha}, \citenamefont {Park},\ and\ \citenamefont
  {Jeong}}]{Lee2020}%
  \BibitemOpen
  \bibfield  {author} {\bibinfo {author} {\bibfnamefont {S.}~\bibnamefont
  {Lee}}, \bibinfo {author} {\bibfnamefont {M.}~\bibnamefont {Ha}}, \bibinfo
  {author} {\bibfnamefont {J.-M.}\ \bibnamefont {Park}},\ and\ \bibinfo
  {author} {\bibfnamefont {H.}~\bibnamefont {Jeong}},\ }\bibfield  {title}
  {\bibinfo {title} {Finite-time quantum otto engine: Surpassing the
  quasistatic efficiency due to friction},\ }\href
  {https://doi.org/10.1103/PhysRevE.101.022127} {\bibfield  {journal} {\bibinfo
   {journal} {Phys. Rev. E}\ }\textbf {\bibinfo {volume} {101}},\ \bibinfo
  {pages} {022127} (\bibinfo {year} {2020})}\BibitemShut {NoStop}%
\bibitem [{\citenamefont {Geva}\ and\ \citenamefont
  {Kosloff}(1992)}]{Geva1992}%
  \BibitemOpen
  \bibfield  {author} {\bibinfo {author} {\bibfnamefont {E.}~\bibnamefont
  {Geva}}\ and\ \bibinfo {author} {\bibfnamefont {R.}~\bibnamefont {Kosloff}},\
  }\bibfield  {title} {\bibinfo {title} {A quantum-mechanical heat engine
  operating in finite time. a model consisting of spin-1/2 systems as the
  working fluid},\ }\href {https://api.semanticscholar.org/CorpusID:121561116}
  {\bibfield  {journal} {\bibinfo  {journal} {Journal of Chemical Physics}\
  }\textbf {\bibinfo {volume} {96}},\ \bibinfo {pages} {3054} (\bibinfo {year}
  {1992})}\BibitemShut {NoStop}%
\bibitem [{\citenamefont {Saha}\ \emph {et~al.}(2023)\citenamefont {Saha},
  \citenamefont {Ghoshal},\ and\ \citenamefont {Sen}}]{saha2023temperature}%
  \BibitemOpen
  \bibfield  {author} {\bibinfo {author} {\bibfnamefont {D.}~\bibnamefont
  {Saha}}, \bibinfo {author} {\bibfnamefont {A.}~\bibnamefont {Ghoshal}},\ and\
  \bibinfo {author} {\bibfnamefont {U.}~\bibnamefont {Sen}},\ }\href@noop {}
  {\bibinfo {title} {Temperature- and interaction-tweaked efficiency boost of
  finite-time robust quantum otto engines}} (\bibinfo {year} {2023}),\ \Eprint
  {https://arxiv.org/abs/2309.11483} {arXiv:2309.11483 [quant-ph]} \BibitemShut
  {NoStop}%
\bibitem [{\citenamefont {Feldmann}\ and\ \citenamefont
  {Kosloff}(2000)}]{feldmann2000performance}%
  \BibitemOpen
  \bibfield  {author} {\bibinfo {author} {\bibfnamefont {T.}~\bibnamefont
  {Feldmann}}\ and\ \bibinfo {author} {\bibfnamefont {R.}~\bibnamefont
  {Kosloff}},\ }\bibfield  {title} {\bibinfo {title} {Performance of discrete
  heat engines and heat pumps in finite time},\ }\href
  {https://doi.org/https://link.aps.org/doi/10.1103/PhysRevE.61.4774}
  {\bibfield  {journal} {\bibinfo  {journal} {Phys. Rev. E}\ }\textbf {\bibinfo
  {volume} {61}},\ \bibinfo {pages} {4774} (\bibinfo {year}
  {2000})}\BibitemShut {NoStop}%
\bibitem [{\citenamefont {Alicki}(2014)}]{alicki2014quantum}%
  \BibitemOpen
  \bibfield  {author} {\bibinfo {author} {\bibfnamefont {R.}~\bibnamefont
  {Alicki}},\ }\bibfield  {title} {\bibinfo {title} {Quantum thermodynamics: An
  example of two-level quantum machine},\ }\href
  {https://doi.org/https://doi.org/10.1142/S1230161214400022} {\bibfield
  {journal} {\bibinfo  {journal} {Open Systems \& Information Dynamics}\
  }\textbf {\bibinfo {volume} {21}},\ \bibinfo {pages} {1440002} (\bibinfo
  {year} {2014})}\BibitemShut {NoStop}%
\bibitem [{\citenamefont {Kieu}(2004)}]{kieu2004}%
  \BibitemOpen
  \bibfield  {author} {\bibinfo {author} {\bibfnamefont {T.~D.}\ \bibnamefont
  {Kieu}},\ }\bibfield  {title} {\bibinfo {title} {The {S}econd law,
  {M}axwell's demon, and work derivable from quantum heat engines},\ }\href
  {https://doi.org/10.1103/PhysRevLett.93.140403} {\bibfield  {journal}
  {\bibinfo  {journal} {Phys. Rev. Lett.}\ }\textbf {\bibinfo {volume} {93}},\
  \bibinfo {pages} {140403} (\bibinfo {year} {2004})}\BibitemShut {NoStop}%
\bibitem [{\citenamefont {Kieu}(2005)}]{kieu2005quantum}%
  \BibitemOpen
  \bibfield  {author} {\bibinfo {author} {\bibfnamefont {T.~D.}\ \bibnamefont
  {Kieu}},\ }\bibfield  {title} {\bibinfo {title} {Quantum heat engines, the
  second law and maxwell's daemon},\ }\href
  {https://doi.org/https://link.springer.com/article/10.1140/epjd/e2006-00075-5}
  {\bibfield  {journal} {\bibinfo  {journal} {The European Physical Journal
  D-Atomic, Molecular, Optical and Plasma Physics}\ ,\ \bibinfo {pages} {115}}
  (\bibinfo {year} {2005})}\BibitemShut {NoStop}%
\bibitem [{\citenamefont {Henrich}\ \emph {et~al.}(2007)\citenamefont
  {Henrich}, \citenamefont {Rempp},\ and\ \citenamefont {Mahler}}]{Mahler2007}%
  \BibitemOpen
  \bibfield  {author} {\bibinfo {author} {\bibfnamefont {M.}~\bibnamefont
  {Henrich}}, \bibinfo {author} {\bibfnamefont {F.}~\bibnamefont {Rempp}},\
  and\ \bibinfo {author} {\bibfnamefont {G.}~\bibnamefont {Mahler}},\
  }\bibfield  {title} {\bibinfo {title} {Quantum thermodynamic {O}tto machines:
  A spin-system approach},\ }\href
  {https://doi.org/10.1140/epjst/e2007-00371-8} {\bibfield  {journal} {\bibinfo
   {journal} {The European Physical Journal Special Topics}\ }\textbf {\bibinfo
  {volume} {151}},\ \bibinfo {pages} {157} (\bibinfo {year}
  {2007})}\BibitemShut {NoStop}%
\bibitem [{\citenamefont {Beretta}(2012)}]{Beretta_2012}%
  \BibitemOpen
  \bibfield  {author} {\bibinfo {author} {\bibfnamefont {G.~P.}\ \bibnamefont
  {Beretta}},\ }\bibfield  {title} {\bibinfo {title} {Quantum thermodynamic
  {C}arnot and {O}tto-like cycles for a two-level system},\ }\href
  {https://doi.org/10.1209/0295-5075/99/20005} {\bibfield  {journal} {\bibinfo
  {journal} {Europhysics Letters}\ }\textbf {\bibinfo {volume} {99}},\ \bibinfo
  {pages} {20005} (\bibinfo {year} {2012})}\BibitemShut {NoStop}%
\bibitem [{\citenamefont {Ghosh}\ \emph {et~al.}(2018)\citenamefont {Ghosh},
  \citenamefont {Gelbwaser-Klimovsky}, \citenamefont {Niedenzu}, \citenamefont
  {Lvovsky}, \citenamefont {Mazets}, \citenamefont {Scully},\ and\
  \citenamefont {Kurizki}}]{Ghosh_2018}%
  \BibitemOpen
  \bibfield  {author} {\bibinfo {author} {\bibfnamefont {A.}~\bibnamefont
  {Ghosh}}, \bibinfo {author} {\bibfnamefont {D.}~\bibnamefont
  {Gelbwaser-Klimovsky}}, \bibinfo {author} {\bibfnamefont {W.}~\bibnamefont
  {Niedenzu}}, \bibinfo {author} {\bibfnamefont {A.~I.}\ \bibnamefont
  {Lvovsky}}, \bibinfo {author} {\bibfnamefont {I.}~\bibnamefont {Mazets}},
  \bibinfo {author} {\bibfnamefont {M.~O.}\ \bibnamefont {Scully}},\ and\
  \bibinfo {author} {\bibfnamefont {G.}~\bibnamefont {Kurizki}},\ }\bibfield
  {title} {\bibinfo {title} {Two-level masers as heat-to-work converters},\
  }\href {https://doi.org/10.1073/pnas.1805354115} {\bibfield  {journal}
  {\bibinfo  {journal} {Proceedings of the National Academy of Sciences}\
  }\textbf {\bibinfo {volume} {115}},\ \bibinfo {pages} {9941–9944} (\bibinfo
  {year} {2018})}\BibitemShut {NoStop}%
\bibitem [{\citenamefont {Papadatos}(2023)}]{Papadatos_2023}%
  \BibitemOpen
  \bibfield  {author} {\bibinfo {author} {\bibfnamefont {N.}~\bibnamefont
  {Papadatos}},\ }\bibfield  {title} {\bibinfo {title} {Quantum {S}tirling heat
  engine with squeezed thermal reservoir},\ }\href
  {https://doi.org/10.1088/1674-1056/acc7f8} {\bibfield  {journal} {\bibinfo
  {journal} {Chinese Physics B}\ }\textbf {\bibinfo {volume} {32}},\ \bibinfo
  {pages} {100702} (\bibinfo {year} {2023})}\BibitemShut {NoStop}%
\bibitem [{\citenamefont {Quan}\ \emph {et~al.}(2005)\citenamefont {Quan},
  \citenamefont {Zhang},\ and\ \citenamefont {Sun}}]{quan2005}%
  \BibitemOpen
  \bibfield  {author} {\bibinfo {author} {\bibfnamefont {H.~T.}\ \bibnamefont
  {Quan}}, \bibinfo {author} {\bibfnamefont {P.}~\bibnamefont {Zhang}},\ and\
  \bibinfo {author} {\bibfnamefont {C.~P.}\ \bibnamefont {Sun}},\ }\bibfield
  {title} {\bibinfo {title} {Quantum heat engine with multilevel quantum
  systems},\ }\href {https://doi.org/10.1103/PhysRevE.72.056110} {\bibfield
  {journal} {\bibinfo  {journal} {Phys. Rev. E}\ }\textbf {\bibinfo {volume}
  {72}},\ \bibinfo {pages} {056110} (\bibinfo {year} {2005})}\BibitemShut
  {NoStop}%
\bibitem [{\citenamefont {Uzdin}\ and\ \citenamefont
  {Kosloff}(2014)}]{Uzdin_2014}%
  \BibitemOpen
  \bibfield  {author} {\bibinfo {author} {\bibfnamefont {R.}~\bibnamefont
  {Uzdin}}\ and\ \bibinfo {author} {\bibfnamefont {R.}~\bibnamefont
  {Kosloff}},\ }\bibfield  {title} {\bibinfo {title} {The multilevel
  four-stroke swap engine and its environment},\ }\href
  {https://doi.org/10.1088/1367-2630/16/9/095003} {\bibfield  {journal}
  {\bibinfo  {journal} {New Journal of Physics}\ }\textbf {\bibinfo {volume}
  {16}},\ \bibinfo {pages} {095003} (\bibinfo {year} {2014})}\BibitemShut
  {NoStop}%
\bibitem [{\citenamefont {de~Oliveira}\ and\ \citenamefont
  {Jonathan}(2021)}]{TDEROliveira2020}%
  \BibitemOpen
  \bibfield  {author} {\bibinfo {author} {\bibfnamefont {T.~R.}\ \bibnamefont
  {de~Oliveira}}\ and\ \bibinfo {author} {\bibfnamefont {D.}~\bibnamefont
  {Jonathan}},\ }\bibfield  {title} {\bibinfo {title} {Efficiency gain and
  bidirectional operation of quantum engines with decoupled internal levels},\
  }\href {https://doi.org/10.1103/PhysRevE.104.044133} {\bibfield  {journal}
  {\bibinfo  {journal} {Phys. Rev. E}\ }\textbf {\bibinfo {volume} {104}},\
  \bibinfo {pages} {044133} (\bibinfo {year} {2021})}\BibitemShut {NoStop}%
\bibitem [{\citenamefont {Simmons}\ \emph {et~al.}(2023)\citenamefont
  {Simmons}, \citenamefont {Sajjad}, \citenamefont {Keithley}, \citenamefont
  {Mas}, \citenamefont {Tanlimco}, \citenamefont {Nolasco-Martinez},
  \citenamefont {Bai}, \citenamefont {Fredrickson},\ and\ \citenamefont
  {Weld}}]{Simmons2023}%
  \BibitemOpen
  \bibfield  {author} {\bibinfo {author} {\bibfnamefont {E.~Q.}\ \bibnamefont
  {Simmons}}, \bibinfo {author} {\bibfnamefont {R.}~\bibnamefont {Sajjad}},
  \bibinfo {author} {\bibfnamefont {K.}~\bibnamefont {Keithley}}, \bibinfo
  {author} {\bibfnamefont {H.}~\bibnamefont {Mas}}, \bibinfo {author}
  {\bibfnamefont {J.~L.}\ \bibnamefont {Tanlimco}}, \bibinfo {author}
  {\bibfnamefont {E.}~\bibnamefont {Nolasco-Martinez}}, \bibinfo {author}
  {\bibfnamefont {Y.}~\bibnamefont {Bai}}, \bibinfo {author} {\bibfnamefont
  {G.~H.}\ \bibnamefont {Fredrickson}},\ and\ \bibinfo {author} {\bibfnamefont
  {D.~M.}\ \bibnamefont {Weld}},\ }\bibfield  {title} {\bibinfo {title}
  {Thermodynamic engine with a quantum degenerate working fluid},\ }\href
  {https://doi.org/10.1103/PhysRevResearch.5.L042009} {\bibfield  {journal}
  {\bibinfo  {journal} {Phys. Rev. Res.}\ }\textbf {\bibinfo {volume} {5}},\
  \bibinfo {pages} {L042009} (\bibinfo {year} {2023})}\BibitemShut {NoStop}%
\bibitem [{\citenamefont {Jaseem}\ \emph {et~al.}(2023)\citenamefont {Jaseem},
  \citenamefont {Vinjanampathy},\ and\ \citenamefont {Mukherjee}}]{Jaseem2023}%
  \BibitemOpen
  \bibfield  {author} {\bibinfo {author} {\bibfnamefont {N.}~\bibnamefont
  {Jaseem}}, \bibinfo {author} {\bibfnamefont {S.}~\bibnamefont
  {Vinjanampathy}},\ and\ \bibinfo {author} {\bibfnamefont {V.}~\bibnamefont
  {Mukherjee}},\ }\bibfield  {title} {\bibinfo {title} {Quadratic enhancement
  in the reliability of collective quantum engines},\ }\href
  {https://doi.org/10.1103/PhysRevA.107.L040202} {\bibfield  {journal}
  {\bibinfo  {journal} {Phys. Rev. A}\ }\textbf {\bibinfo {volume} {107}},\
  \bibinfo {pages} {L040202} (\bibinfo {year} {2023})}\BibitemShut {NoStop}%
\bibitem [{\citenamefont {Williamson}\ and\ \citenamefont
  {Davis}(2024)}]{Williamson2024}%
  \BibitemOpen
  \bibfield  {author} {\bibinfo {author} {\bibfnamefont {L.~A.}\ \bibnamefont
  {Williamson}}\ and\ \bibinfo {author} {\bibfnamefont {M.~J.}\ \bibnamefont
  {Davis}},\ }\bibfield  {title} {\bibinfo {title} {Many-body enhancement in a
  spin-chain quantum heat engine},\ }\href
  {https://doi.org/10.1103/PhysRevB.109.024310} {\bibfield  {journal} {\bibinfo
   {journal} {Phys. Rev. B}\ }\textbf {\bibinfo {volume} {109}},\ \bibinfo
  {pages} {024310} (\bibinfo {year} {2024})}\BibitemShut {NoStop}%
\bibitem [{\citenamefont {Marshall}\ \emph {et~al.}(2011)\citenamefont
  {Marshall}, \citenamefont {Olkin},\ and\ \citenamefont
  {Arnold}}]{Marshallmajorizationbook}%
  \BibitemOpen
  \bibfield  {author} {\bibinfo {author} {\bibfnamefont {A.~W.}\ \bibnamefont
  {Marshall}}, \bibinfo {author} {\bibfnamefont {I.}~\bibnamefont {Olkin}},\
  and\ \bibinfo {author} {\bibfnamefont {B.~C.}\ \bibnamefont {Arnold}},\
  }\href {https://doi.org/https://doi.org/10.1007/978-0-387-68276-1} {\emph
  {\bibinfo {title} {Inequalities: Theory of Majorization and Its
  Applications}}}\ (\bibinfo  {publisher} {Springer Series in Statistics,
  Springer, New York},\ \bibinfo {year} {2011})\BibitemShut {NoStop}%
\bibitem [{\citenamefont {Sagawa}(2020)}]{TSagawa}%
  \BibitemOpen
  \bibfield  {author} {\bibinfo {author} {\bibfnamefont {T.}~\bibnamefont
  {Sagawa}},\ }\href
  {https://doi.org/https://doi.org/10.1007/978-981-16-6644-5} {\emph {\bibinfo
  {title} {Entropy, Divergence, and Majorization in Classical and Quantum
  Thermodynamics}}}\ (\bibinfo  {publisher} {Springer Briefs in Mathematical
  Physics, Springer Singapore},\ \bibinfo {year} {2020})\BibitemShut {NoStop}%
\bibitem [{\citenamefont {Bhatia}(1996)}]{Bhatia1996MatrixA}%
  \BibitemOpen
  \bibfield  {author} {\bibinfo {author} {\bibfnamefont {R.}~\bibnamefont
  {Bhatia}},\ }\href
  {https://doi.org/https://doi.org/10.1007/978-1-4612-0653-8} {\emph {\bibinfo
  {title} {Matrix Analysis}}}\ (\bibinfo  {publisher} {Springer New York},\
  \bibinfo {year} {1996})\BibitemShut {NoStop}%
\bibitem [{\citenamefont {Buscemi}\ and\ \citenamefont
  {Gour}(2017)}]{Buscemi2017}%
  \BibitemOpen
  \bibfield  {author} {\bibinfo {author} {\bibfnamefont {F.}~\bibnamefont
  {Buscemi}}\ and\ \bibinfo {author} {\bibfnamefont {G.}~\bibnamefont {Gour}},\
  }\bibfield  {title} {\bibinfo {title} {Quantum relative {L}orenz curves},\
  }\href {https://doi.org/10.1103/PhysRevA.95.012110} {\bibfield  {journal}
  {\bibinfo  {journal} {Phys. Rev. A}\ }\textbf {\bibinfo {volume} {95}},\
  \bibinfo {pages} {012110} (\bibinfo {year} {2017})}\BibitemShut {NoStop}%
\bibitem [{\citenamefont {Joe}(1990)}]{Joe1990MajorizationAD}%
  \BibitemOpen
  \bibfield  {author} {\bibinfo {author} {\bibfnamefont {H.}~\bibnamefont
  {Joe}},\ }\bibfield  {title} {\bibinfo {title} {Majorization and
  divergence},\ }\href
  {https://doi.org/https://doi.org/10.1016/0022-247X(90)90002-W} {\bibfield
  {journal} {\bibinfo  {journal} {Journal of Mathematical Analysis and
  Applications}\ }\textbf {\bibinfo {volume} {148}},\ \bibinfo {pages} {287}
  (\bibinfo {year} {1990})}\BibitemShut {NoStop}%
\bibitem [{\citenamefont {Shiraishi}(2020)}]{2020}%
  \BibitemOpen
  \bibfield  {author} {\bibinfo {author} {\bibfnamefont {N.}~\bibnamefont
  {Shiraishi}},\ }\bibfield  {title} {\bibinfo {title} {Two constructive proofs
  on d-majorization and thermo-majorization},\ }\href
  {https://doi.org/10.1088/1751-8121/abb041} {\bibfield  {journal} {\bibinfo
  {journal} {Journal of Physics A: Mathematical and Theoretical}\ }\textbf
  {\bibinfo {volume} {53}},\ \bibinfo {pages} {425301} (\bibinfo {year}
  {2020})}\BibitemShut {NoStop}%
\bibitem [{\citenamefont {Egloff}\ \emph {et~al.}(2015)\citenamefont {Egloff},
  \citenamefont {Dahlsten}, \citenamefont {Renner},\ and\ \citenamefont
  {Vedral}}]{vedral2015}%
  \BibitemOpen
  \bibfield  {author} {\bibinfo {author} {\bibfnamefont {D.}~\bibnamefont
  {Egloff}}, \bibinfo {author} {\bibfnamefont {O.~C.~O.}\ \bibnamefont
  {Dahlsten}}, \bibinfo {author} {\bibfnamefont {R.}~\bibnamefont {Renner}},\
  and\ \bibinfo {author} {\bibfnamefont {V.}~\bibnamefont {Vedral}},\
  }\bibfield  {title} {\bibinfo {title} {A measure of majorization emerging
  from single-shot statistical mechanics},\ }\href
  {https://doi.org/10.1088/1367-2630/17/7/073001} {\bibfield  {journal}
  {\bibinfo  {journal} {New Journal of Physics}\ }\textbf {\bibinfo {volume}
  {17}},\ \bibinfo {pages} {073001} (\bibinfo {year} {2015})}\BibitemShut
  {NoStop}%
\bibitem [{\citenamefont {Renes}(2016)}]{2016}%
  \BibitemOpen
  \bibfield  {author} {\bibinfo {author} {\bibfnamefont {J.~M.}\ \bibnamefont
  {Renes}},\ }\bibfield  {title} {\bibinfo {title} {Relative submajorization
  and its use in quantum resource theories},\ }\href
  {https://doi.org/10.1063/1.4972295} {\bibfield  {journal} {\bibinfo
  {journal} {Journal of Mathematical Physics}\ }\textbf {\bibinfo {volume}
  {57}},\ \bibinfo {pages} {122202} (\bibinfo {year} {2016})}\BibitemShut
  {NoStop}%
\bibitem [{\citenamefont {Ruch}\ \emph {et~al.}(1980)\citenamefont {Ruch},
  \citenamefont {Schranner},\ and\ \citenamefont {Seligman}}]{RUCH1980222}%
  \BibitemOpen
  \bibfield  {author} {\bibinfo {author} {\bibfnamefont {E.}~\bibnamefont
  {Ruch}}, \bibinfo {author} {\bibfnamefont {R.}~\bibnamefont {Schranner}},\
  and\ \bibinfo {author} {\bibfnamefont {T.~H.}\ \bibnamefont {Seligman}},\
  }\bibfield  {title} {\bibinfo {title} {Generalization of a theorem by
  {H}ardy, {L}ittlewood, and {P}ólya},\ }\href
  {https://doi.org/10.1016/0022-247X(80)90075-X} {\bibfield  {journal}
  {\bibinfo  {journal} {Journal of Mathematical Analysis and Applications}\
  }\textbf {\bibinfo {volume} {76}},\ \bibinfo {pages} {222–229} (\bibinfo
  {year} {1980})}\BibitemShut {NoStop}%
\bibitem [{\citenamefont {Nielsen}\ and\ \citenamefont
  {Vidal}(2001)}]{10.5555/2011326.2011331}%
  \BibitemOpen
  \bibfield  {author} {\bibinfo {author} {\bibfnamefont {M.~A.}\ \bibnamefont
  {Nielsen}}\ and\ \bibinfo {author} {\bibfnamefont {G.}~\bibnamefont
  {Vidal}},\ }\bibfield  {title} {\bibinfo {title} {Majorization and the
  interconversion of bipartite states},\ }\href
  {https://doi.org/https://dl.acm.org/doi/abs/10.5555/2011326.2011331}
  {\bibfield  {journal} {\bibinfo  {journal} {Quantum Info. Comput.}\ }\textbf
  {\bibinfo {volume} {1}},\ \bibinfo {pages} {76–93} (\bibinfo {year}
  {2001})}\BibitemShut {NoStop}%
\bibitem [{\citenamefont {Nielsen}(1999)}]{PhysRevLett.83.436}%
  \BibitemOpen
  \bibfield  {author} {\bibinfo {author} {\bibfnamefont {M.~A.}\ \bibnamefont
  {Nielsen}},\ }\bibfield  {title} {\bibinfo {title} {Conditions for a class of
  entanglement transformations},\ }\href
  {https://doi.org/10.1103/PhysRevLett.83.436} {\bibfield  {journal} {\bibinfo
  {journal} {Phys. Rev. Lett.}\ }\textbf {\bibinfo {volume} {83}},\ \bibinfo
  {pages} {436} (\bibinfo {year} {1999})}\BibitemShut {NoStop}%
\bibitem [{\citenamefont {Du}\ \emph {et~al.}(2015)\citenamefont {Du},
  \citenamefont {Bai},\ and\ \citenamefont {Guo}}]{PhysRevA.91.052120}%
  \BibitemOpen
  \bibfield  {author} {\bibinfo {author} {\bibfnamefont {S.}~\bibnamefont
  {Du}}, \bibinfo {author} {\bibfnamefont {Z.}~\bibnamefont {Bai}},\ and\
  \bibinfo {author} {\bibfnamefont {Y.}~\bibnamefont {Guo}},\ }\bibfield
  {title} {\bibinfo {title} {Conditions for coherence transformations under
  incoherent operations},\ }\href {https://doi.org/10.1103/PhysRevA.91.052120}
  {\bibfield  {journal} {\bibinfo  {journal} {Phys. Rev. A}\ }\textbf {\bibinfo
  {volume} {91}},\ \bibinfo {pages} {052120} (\bibinfo {year}
  {2015})}\BibitemShut {NoStop}%
\bibitem [{\citenamefont {Jonathan}\ and\ \citenamefont
  {Plenio}(1999)}]{PhysRevLett.83.3566}%
  \BibitemOpen
  \bibfield  {author} {\bibinfo {author} {\bibfnamefont {D.}~\bibnamefont
  {Jonathan}}\ and\ \bibinfo {author} {\bibfnamefont {M.~B.}\ \bibnamefont
  {Plenio}},\ }\bibfield  {title} {\bibinfo {title} {Entanglement-assisted
  local manipulation of pure quantum states},\ }\href
  {https://doi.org/10.1103/PhysRevLett.83.3566} {\bibfield  {journal} {\bibinfo
   {journal} {Phys. Rev. Lett.}\ }\textbf {\bibinfo {volume} {83}},\ \bibinfo
  {pages} {3566} (\bibinfo {year} {1999})}\BibitemShut {NoStop}%
\bibitem [{\citenamefont {Horodecki}\ \emph {et~al.}(2003)\citenamefont
  {Horodecki}, \citenamefont {Horodecki},\ and\ \citenamefont
  {Oppenheim}}]{Horodecki2003}%
  \BibitemOpen
  \bibfield  {author} {\bibinfo {author} {\bibfnamefont {M.}~\bibnamefont
  {Horodecki}}, \bibinfo {author} {\bibfnamefont {P.}~\bibnamefont
  {Horodecki}},\ and\ \bibinfo {author} {\bibfnamefont {J.}~\bibnamefont
  {Oppenheim}},\ }\bibfield  {title} {\bibinfo {title} {Reversible
  transformations from pure to mixed states and the unique measure of
  information},\ }\href {https://doi.org/10.1103/PhysRevA.67.062104} {\bibfield
   {journal} {\bibinfo  {journal} {Phys. Rev. A}\ }\textbf {\bibinfo {volume}
  {67}},\ \bibinfo {pages} {062104} (\bibinfo {year} {2003})}\BibitemShut
  {NoStop}%
\bibitem [{\citenamefont {Singh}\ \emph {et~al.}(2021)\citenamefont {Singh},
  \citenamefont {Das},\ and\ \citenamefont {Cerf}}]{Uttam2021}%
  \BibitemOpen
  \bibfield  {author} {\bibinfo {author} {\bibfnamefont {U.}~\bibnamefont
  {Singh}}, \bibinfo {author} {\bibfnamefont {S.}~\bibnamefont {Das}},\ and\
  \bibinfo {author} {\bibfnamefont {N.~J.}\ \bibnamefont {Cerf}},\ }\bibfield
  {title} {\bibinfo {title} {Partial order on passive states and {H}offman
  majorization in quantum thermodynamics},\ }\href
  {https://doi.org/10.1103/PhysRevResearch.3.033091} {\bibfield  {journal}
  {\bibinfo  {journal} {Phys. Rev. Res.}\ }\textbf {\bibinfo {volume} {3}},\
  \bibinfo {pages} {033091} (\bibinfo {year} {2021})}\BibitemShut {NoStop}%
\bibitem [{\citenamefont {Lostaglio}\ and\ \citenamefont
  {Korzekwa}(2022)}]{lostaglio2022}%
  \BibitemOpen
  \bibfield  {author} {\bibinfo {author} {\bibfnamefont {M.}~\bibnamefont
  {Lostaglio}}\ and\ \bibinfo {author} {\bibfnamefont {K.}~\bibnamefont
  {Korzekwa}},\ }\bibfield  {title} {\bibinfo {title} {Continuous
  thermomajorization and a complete set of laws for markovian thermal
  processes},\ }\href {https://doi.org/10.1103/PhysRevA.106.012426} {\bibfield
  {journal} {\bibinfo  {journal} {Phys. Rev. A}\ }\textbf {\bibinfo {volume}
  {106}},\ \bibinfo {pages} {012426} (\bibinfo {year} {2022})}\BibitemShut
  {NoStop}%
\bibitem [{\citenamefont {Gour}\ \emph {et~al.}(2018)\citenamefont {Gour},
  \citenamefont {Jennings}, \citenamefont {Buscemi}, \citenamefont {Duan},\
  and\ \citenamefont {Marvian}}]{majorizationcomplete}%
  \BibitemOpen
  \bibfield  {author} {\bibinfo {author} {\bibfnamefont {G.}~\bibnamefont
  {Gour}}, \bibinfo {author} {\bibfnamefont {D.}~\bibnamefont {Jennings}},
  \bibinfo {author} {\bibfnamefont {F.}~\bibnamefont {Buscemi}}, \bibinfo
  {author} {\bibfnamefont {R.}~\bibnamefont {Duan}},\ and\ \bibinfo {author}
  {\bibfnamefont {I.}~\bibnamefont {Marvian}},\ }\bibfield  {title} {\bibinfo
  {title} {Quantum majorization and a complete set of entropic conditions for
  quantum thermodynamics},\ }\href {https://doi.org/10.1038/s41467-018-06261-7}
  {\bibfield  {journal} {\bibinfo  {journal} {Nature Communications}\ }\textbf
  {\bibinfo {volume} {9}},\ \bibinfo {pages} {5352} (\bibinfo {year}
  {2018})}\BibitemShut {NoStop}%
\bibitem [{\citenamefont {Rethinasamy}\ and\ \citenamefont
  {Wilde}(2020)}]{PhysRevResearch.2.033455}%
  \BibitemOpen
  \bibfield  {author} {\bibinfo {author} {\bibfnamefont {S.}~\bibnamefont
  {Rethinasamy}}\ and\ \bibinfo {author} {\bibfnamefont {M.~M.}\ \bibnamefont
  {Wilde}},\ }\bibfield  {title} {\bibinfo {title} {Relative entropy and
  catalytic relative majorization},\ }\href
  {https://doi.org/10.1103/PhysRevResearch.2.033455} {\bibfield  {journal}
  {\bibinfo  {journal} {Phys. Rev. Research}\ }\textbf {\bibinfo {volume}
  {2}},\ \bibinfo {pages} {033455} (\bibinfo {year} {2020})}\BibitemShut
  {NoStop}%
\bibitem [{\citenamefont {Brandão}\ \emph {et~al.}(2015)\citenamefont
  {Brandão}, \citenamefont {Horodecki}, \citenamefont {Ng}, \citenamefont
  {Oppenheim},\ and\ \citenamefont {Wehner}}]{2015}%
  \BibitemOpen
  \bibfield  {author} {\bibinfo {author} {\bibfnamefont {F.}~\bibnamefont
  {Brandão}}, \bibinfo {author} {\bibfnamefont {M.}~\bibnamefont {Horodecki}},
  \bibinfo {author} {\bibfnamefont {N.}~\bibnamefont {Ng}}, \bibinfo {author}
  {\bibfnamefont {J.}~\bibnamefont {Oppenheim}},\ and\ \bibinfo {author}
  {\bibfnamefont {S.}~\bibnamefont {Wehner}},\ }\bibfield  {title} {\bibinfo
  {title} {The second laws of quantum thermodynamics},\ }\href
  {https://doi.org/10.1073/pnas.1411728112} {\bibfield  {journal} {\bibinfo
  {journal} {Proceedings of the National Academy of Sciences}\ }\textbf
  {\bibinfo {volume} {112}},\ \bibinfo {pages} {3275–3279} (\bibinfo {year}
  {2015})}\BibitemShut {NoStop}%
\bibitem [{\citenamefont {Horodecki}\ and\ \citenamefont
  {Oppenheim}(2013)}]{2013}%
  \BibitemOpen
  \bibfield  {author} {\bibinfo {author} {\bibfnamefont {M.}~\bibnamefont
  {Horodecki}}\ and\ \bibinfo {author} {\bibfnamefont {J.}~\bibnamefont
  {Oppenheim}},\ }\bibfield  {title} {\bibinfo {title} {Fundamental limitations
  for quantum and nanoscale thermodynamics},\ }\href
  {https://doi.org/10.1038/ncomms3059} {\bibfield  {journal} {\bibinfo
  {journal} {Nature Communications}\ }\textbf {\bibinfo {volume} {4}},\
  \bibinfo {pages} {2059} (\bibinfo {year} {2013})}\BibitemShut {NoStop}%
\bibitem [{\citenamefont {Junior}\ \emph {et~al.}(2022)\citenamefont {Junior},
  \citenamefont {Czartowski}, \citenamefont {\ifmmode~\dot{Z}\else
  \.{Z}\fi{}yczkowski},\ and\ \citenamefont {Korzekwa}}]{thermalcone_2022}%
  \BibitemOpen
  \bibfield  {author} {\bibinfo {author} {\bibfnamefont {A.~d.~O.}\
  \bibnamefont {Junior}}, \bibinfo {author} {\bibfnamefont {J.}~\bibnamefont
  {Czartowski}}, \bibinfo {author} {\bibfnamefont {K.}~\bibnamefont
  {\ifmmode~\dot{Z}\else \.{Z}\fi{}yczkowski}},\ and\ \bibinfo {author}
  {\bibfnamefont {K.}~\bibnamefont {Korzekwa}},\ }\bibfield  {title} {\bibinfo
  {title} {Geometric structure of thermal cones},\ }\href
  {https://doi.org/10.1103/PhysRevE.106.064109} {\bibfield  {journal} {\bibinfo
   {journal} {Phys. Rev. E}\ }\textbf {\bibinfo {volume} {106}},\ \bibinfo
  {pages} {064109} (\bibinfo {year} {2022})}\BibitemShut {NoStop}%
\bibitem [{\citenamefont {Bosyk}\ \emph {et~al.}(2019)\citenamefont {Bosyk},
  \citenamefont {Bellomo}, \citenamefont {Holik}, \citenamefont {Freytes},\
  and\ \citenamefont {Sergioli}}]{Bosyk_2019}%
  \BibitemOpen
  \bibfield  {author} {\bibinfo {author} {\bibfnamefont {G.~M.}\ \bibnamefont
  {Bosyk}}, \bibinfo {author} {\bibfnamefont {G.}~\bibnamefont {Bellomo}},
  \bibinfo {author} {\bibfnamefont {F.}~\bibnamefont {Holik}}, \bibinfo
  {author} {\bibfnamefont {H.}~\bibnamefont {Freytes}},\ and\ \bibinfo {author}
  {\bibfnamefont {G.}~\bibnamefont {Sergioli}},\ }\bibfield  {title} {\bibinfo
  {title} {Optimal common resource in majorization-based resource theories},\
  }\href {https://doi.org/10.1088/1367-2630/ab3734} {\bibfield  {journal}
  {\bibinfo  {journal} {New Journal of Physics}\ }\textbf {\bibinfo {volume}
  {21}},\ \bibinfo {pages} {083028} (\bibinfo {year} {2019})}\BibitemShut
  {NoStop}%
\bibitem [{\citenamefont {Alimuddin}\ \emph
  {et~al.}(2020{\natexlab{a}})\citenamefont {Alimuddin}, \citenamefont {Guha},\
  and\ \citenamefont {Parashar}}]{Mir2020}%
  \BibitemOpen
  \bibfield  {author} {\bibinfo {author} {\bibfnamefont {M.}~\bibnamefont
  {Alimuddin}}, \bibinfo {author} {\bibfnamefont {T.}~\bibnamefont {Guha}},\
  and\ \bibinfo {author} {\bibfnamefont {P.}~\bibnamefont {Parashar}},\
  }\bibfield  {title} {\bibinfo {title} {Independence of work and entropy for
  equal-energetic finite quantum systems: Passive-state energy as an
  entanglement quantifier},\ }\href
  {https://doi.org/10.1103/PhysRevE.102.012145} {\bibfield  {journal} {\bibinfo
   {journal} {Phys. Rev. E}\ }\textbf {\bibinfo {volume} {102}},\ \bibinfo
  {pages} {012145} (\bibinfo {year} {2020}{\natexlab{a}})}\BibitemShut
  {NoStop}%
\bibitem [{\citenamefont {Alimuddin}\ \emph {et~al.}(2019)\citenamefont
  {Alimuddin}, \citenamefont {Guha},\ and\ \citenamefont {Parashar}}]{Mir2019}%
  \BibitemOpen
  \bibfield  {author} {\bibinfo {author} {\bibfnamefont {M.}~\bibnamefont
  {Alimuddin}}, \bibinfo {author} {\bibfnamefont {T.}~\bibnamefont {Guha}},\
  and\ \bibinfo {author} {\bibfnamefont {P.}~\bibnamefont {Parashar}},\
  }\bibfield  {title} {\bibinfo {title} {Bound on ergotropic gap for bipartite
  separable states},\ }\href {https://doi.org/10.1103/PhysRevA.99.052320}
  {\bibfield  {journal} {\bibinfo  {journal} {Phys. Rev. A}\ }\textbf {\bibinfo
  {volume} {99}},\ \bibinfo {pages} {052320} (\bibinfo {year}
  {2019})}\BibitemShut {NoStop}%
\bibitem [{\citenamefont {Joshi}\ \emph {et~al.}(2024)\citenamefont {Joshi},
  \citenamefont {Alimuddin}, \citenamefont {Mahesh},\ and\ \citenamefont
  {Banik}}]{Joshi2024}%
  \BibitemOpen
  \bibfield  {author} {\bibinfo {author} {\bibfnamefont {J.}~\bibnamefont
  {Joshi}}, \bibinfo {author} {\bibfnamefont {M.}~\bibnamefont {Alimuddin}},
  \bibinfo {author} {\bibfnamefont {T.~S.}\ \bibnamefont {Mahesh}},\ and\
  \bibinfo {author} {\bibfnamefont {M.}~\bibnamefont {Banik}},\ }\bibfield
  {title} {\bibinfo {title} {Experimental verification of many-body
  entanglement using thermodynamic quantities},\ }\href
  {https://doi.org/10.1103/PhysRevA.109.L020403} {\bibfield  {journal}
  {\bibinfo  {journal} {Phys. Rev. A}\ }\textbf {\bibinfo {volume} {109}},\
  \bibinfo {pages} {L020403} (\bibinfo {year} {2024})}\BibitemShut {NoStop}%
\bibitem [{\citenamefont {Puliyil}\ \emph {et~al.}(2022)\citenamefont
  {Puliyil}, \citenamefont {Banik},\ and\ \citenamefont
  {Alimuddin}}]{puliyil2022}%
  \BibitemOpen
  \bibfield  {author} {\bibinfo {author} {\bibfnamefont {S.}~\bibnamefont
  {Puliyil}}, \bibinfo {author} {\bibfnamefont {M.}~\bibnamefont {Banik}},\
  and\ \bibinfo {author} {\bibfnamefont {M.}~\bibnamefont {Alimuddin}},\
  }\bibfield  {title} {\bibinfo {title} {Thermodynamic signatures of genuinely
  multipartite entanglement},\ }\href
  {https://doi.org/10.1103/PhysRevLett.129.070601} {\bibfield  {journal}
  {\bibinfo  {journal} {Phys. Rev. Lett.}\ }\textbf {\bibinfo {volume} {129}},\
  \bibinfo {pages} {070601} (\bibinfo {year} {2022})}\BibitemShut {NoStop}%
\bibitem [{\citenamefont {Mannalath}\ and\ \citenamefont
  {Pathak}(2023)}]{vaishak2023}%
  \BibitemOpen
  \bibfield  {author} {\bibinfo {author} {\bibfnamefont {V.}~\bibnamefont
  {Mannalath}}\ and\ \bibinfo {author} {\bibfnamefont {A.}~\bibnamefont
  {Pathak}},\ }\bibfield  {title} {\bibinfo {title} {Multiparty entanglement
  routing in quantum networks},\ }\href
  {https://doi.org/10.1103/PhysRevA.108.062614} {\bibfield  {journal} {\bibinfo
   {journal} {Phys. Rev. A}\ }\textbf {\bibinfo {volume} {108}},\ \bibinfo
  {pages} {062614} (\bibinfo {year} {2023})}\BibitemShut {NoStop}%
\bibitem [{\citenamefont {Alimuddin}\ \emph
  {et~al.}(2020{\natexlab{b}})\citenamefont {Alimuddin}, \citenamefont {Guha},\
  and\ \citenamefont {Parashar}}]{MirAli2019}%
  \BibitemOpen
  \bibfield  {author} {\bibinfo {author} {\bibfnamefont {M.}~\bibnamefont
  {Alimuddin}}, \bibinfo {author} {\bibfnamefont {T.}~\bibnamefont {Guha}},\
  and\ \bibinfo {author} {\bibfnamefont {P.}~\bibnamefont {Parashar}},\
  }\bibfield  {title} {\bibinfo {title} {Structure of passive states and its
  implication in charging quantum batteries},\ }\href
  {https://doi.org/10.1103/PhysRevE.102.022106} {\bibfield  {journal} {\bibinfo
   {journal} {Phys. Rev. E}\ }\textbf {\bibinfo {volume} {102}},\ \bibinfo
  {pages} {022106} (\bibinfo {year} {2020}{\natexlab{b}})}\BibitemShut
  {NoStop}%
\bibitem [{\citenamefont {Fedorov}\ \emph {et~al.}(2012)\citenamefont
  {Fedorov}, \citenamefont {Steffen}, \citenamefont {Baur}, \citenamefont
  {da~Silva},\ and\ \citenamefont {Wallraff}}]{Fedorov2012}%
  \BibitemOpen
  \bibfield  {author} {\bibinfo {author} {\bibfnamefont {A.}~\bibnamefont
  {Fedorov}}, \bibinfo {author} {\bibfnamefont {L.}~\bibnamefont {Steffen}},
  \bibinfo {author} {\bibfnamefont {M.}~\bibnamefont {Baur}}, \bibinfo {author}
  {\bibfnamefont {M.~P.}\ \bibnamefont {da~Silva}},\ and\ \bibinfo {author}
  {\bibfnamefont {A.}~\bibnamefont {Wallraff}},\ }\bibfield  {title} {\bibinfo
  {title} {Implementation of a {T}offoli gate with superconducting circuits},\
  }\href {https://doi.org/10.1038/nature10713} {\bibfield  {journal} {\bibinfo
  {journal} {Nature}\ }\textbf {\bibinfo {volume} {481}},\ \bibinfo {pages}
  {170} (\bibinfo {year} {2012})}\BibitemShut {NoStop}%
\bibitem [{\citenamefont {Yurtalan}\ \emph {et~al.}(2020)\citenamefont
  {Yurtalan}, \citenamefont {Shi}, \citenamefont {Kononenko}, \citenamefont
  {Lupascu},\ and\ \citenamefont {Ashhab}}]{Yurtalan2020}%
  \BibitemOpen
  \bibfield  {author} {\bibinfo {author} {\bibfnamefont {M.~A.}\ \bibnamefont
  {Yurtalan}}, \bibinfo {author} {\bibfnamefont {J.}~\bibnamefont {Shi}},
  \bibinfo {author} {\bibfnamefont {M.}~\bibnamefont {Kononenko}}, \bibinfo
  {author} {\bibfnamefont {A.}~\bibnamefont {Lupascu}},\ and\ \bibinfo {author}
  {\bibfnamefont {S.}~\bibnamefont {Ashhab}},\ }\bibfield  {title} {\bibinfo
  {title} {Implementation of a {W}alsh-{H}adamard gate in a superconducting
  qutrit},\ }\href {https://doi.org/10.1103/PhysRevLett.125.180504} {\bibfield
  {journal} {\bibinfo  {journal} {Phys. Rev. Lett.}\ }\textbf {\bibinfo
  {volume} {125}},\ \bibinfo {pages} {180504} (\bibinfo {year}
  {2020})}\BibitemShut {NoStop}%
\bibitem [{\citenamefont {Morvan}\ \emph {et~al.}(2021)\citenamefont {Morvan},
  \citenamefont {Ramasesh}, \citenamefont {Blok}, \citenamefont {Kreikebaum},
  \citenamefont {O'Brien}, \citenamefont {Chen}, \citenamefont {Mitchell},
  \citenamefont {Naik}, \citenamefont {Santiago},\ and\ \citenamefont
  {Siddiqi}}]{Morvan2021}%
  \BibitemOpen
  \bibfield  {author} {\bibinfo {author} {\bibfnamefont {A.}~\bibnamefont
  {Morvan}}, \bibinfo {author} {\bibfnamefont {V.~V.}\ \bibnamefont
  {Ramasesh}}, \bibinfo {author} {\bibfnamefont {M.~S.}\ \bibnamefont {Blok}},
  \bibinfo {author} {\bibfnamefont {J.~M.}\ \bibnamefont {Kreikebaum}},
  \bibinfo {author} {\bibfnamefont {K.}~\bibnamefont {O'Brien}}, \bibinfo
  {author} {\bibfnamefont {L.}~\bibnamefont {Chen}}, \bibinfo {author}
  {\bibfnamefont {B.~K.}\ \bibnamefont {Mitchell}}, \bibinfo {author}
  {\bibfnamefont {R.~K.}\ \bibnamefont {Naik}}, \bibinfo {author}
  {\bibfnamefont {D.~I.}\ \bibnamefont {Santiago}},\ and\ \bibinfo {author}
  {\bibfnamefont {I.}~\bibnamefont {Siddiqi}},\ }\bibfield  {title} {\bibinfo
  {title} {Qutrit randomized benchmarking},\ }\href
  {https://doi.org/10.1103/PhysRevLett.126.210504} {\bibfield  {journal}
  {\bibinfo  {journal} {Phys. Rev. Lett.}\ }\textbf {\bibinfo {volume} {126}},\
  \bibinfo {pages} {210504} (\bibinfo {year} {2021})}\BibitemShut {NoStop}%
\bibitem [{\citenamefont {Díaz}\ and\ \citenamefont
  {Sánchez}(2021)}]{Diaz2021}%
  \BibitemOpen
  \bibfield  {author} {\bibinfo {author} {\bibfnamefont {I.}~\bibnamefont
  {Díaz}}\ and\ \bibinfo {author} {\bibfnamefont {R.}~\bibnamefont
  {Sánchez}},\ }\bibfield  {title} {\bibinfo {title} {The qutrit as a heat
  diode and circulator},\ }\href {https://doi.org/10.1088/1367-2630/ac4211}
  {\bibfield  {journal} {\bibinfo  {journal} {New Journal of Physics}\ }\textbf
  {\bibinfo {volume} {23}},\ \bibinfo {pages} {125006} (\bibinfo {year}
  {2021})}\BibitemShut {NoStop}%
\bibitem [{\citenamefont {Thomas}\ \emph {et~al.}(2020)\citenamefont {Thomas},
  \citenamefont {Gubaydullin}, \citenamefont {Golubev},\ and\ \citenamefont
  {Pekola}}]{George2020}%
  \BibitemOpen
  \bibfield  {author} {\bibinfo {author} {\bibfnamefont {G.}~\bibnamefont
  {Thomas}}, \bibinfo {author} {\bibfnamefont {A.}~\bibnamefont {Gubaydullin}},
  \bibinfo {author} {\bibfnamefont {D.~S.}\ \bibnamefont {Golubev}},\ and\
  \bibinfo {author} {\bibfnamefont {J.~P.}\ \bibnamefont {Pekola}},\ }\bibfield
   {title} {\bibinfo {title} {Thermally pumped on-chip maser},\ }\href
  {https://doi.org/10.1103/PhysRevB.102.104503} {\bibfield  {journal} {\bibinfo
   {journal} {Phys. Rev. B}\ }\textbf {\bibinfo {volume} {102}},\ \bibinfo
  {pages} {104503} (\bibinfo {year} {2020})}\BibitemShut {NoStop}%
\bibitem [{\citenamefont {Gelbwaser-Klimovsky}\ \emph
  {et~al.}(2015)\citenamefont {Gelbwaser-Klimovsky}, \citenamefont {Niedenzu},
  \citenamefont {Brumer},\ and\ \citenamefont {Kurizki}}]{Kurizki2015}%
  \BibitemOpen
  \bibfield  {author} {\bibinfo {author} {\bibfnamefont {D.}~\bibnamefont
  {Gelbwaser-Klimovsky}}, \bibinfo {author} {\bibfnamefont {W.}~\bibnamefont
  {Niedenzu}}, \bibinfo {author} {\bibfnamefont {P.}~\bibnamefont {Brumer}},\
  and\ \bibinfo {author} {\bibfnamefont {G.}~\bibnamefont {Kurizki}},\
  }\bibfield  {title} {\bibinfo {title} {Power enhancement of heat engines via
  correlated thermalization in a three-level ``working fluid''},\ }\href
  {https://doi.org/10.1038/srep14413} {\bibfield  {journal} {\bibinfo
  {journal} {Scientific Reports}\ }\textbf {\bibinfo {volume} {5}},\ \bibinfo
  {pages} {14413} (\bibinfo {year} {2015})}\BibitemShut {NoStop}%
\bibitem [{\citenamefont {Mehta}\ and\ \citenamefont {Johal}(2017)}]{Venu2017}%
  \BibitemOpen
  \bibfield  {author} {\bibinfo {author} {\bibfnamefont {V.}~\bibnamefont
  {Mehta}}\ and\ \bibinfo {author} {\bibfnamefont {R.~S.}\ \bibnamefont
  {Johal}},\ }\bibfield  {title} {\bibinfo {title} {Quantum {O}tto engine with
  exchange coupling in the presence of level degeneracy},\ }\href
  {https://doi.org/10.1103/PhysRevE.96.032110} {\bibfield  {journal} {\bibinfo
  {journal} {Phys. Rev. E}\ }\textbf {\bibinfo {volume} {96}},\ \bibinfo
  {pages} {032110} (\bibinfo {year} {2017})}\BibitemShut {NoStop}%
\bibitem [{\citenamefont {Anka}\ \emph {et~al.}(2021)\citenamefont {Anka},
  \citenamefont {de~Oliveira},\ and\ \citenamefont {Jonathan}}]{anka2021}%
  \BibitemOpen
  \bibfield  {author} {\bibinfo {author} {\bibfnamefont {M.~F.}\ \bibnamefont
  {Anka}}, \bibinfo {author} {\bibfnamefont {T.~R.}\ \bibnamefont
  {de~Oliveira}},\ and\ \bibinfo {author} {\bibfnamefont {D.}~\bibnamefont
  {Jonathan}},\ }\bibfield  {title} {\bibinfo {title} {Measurement-based
  quantum heat engine in a multilevel system},\ }\href
  {https://doi.org/10.1103/PhysRevE.104.054128} {\bibfield  {journal} {\bibinfo
   {journal} {Phys. Rev. E}\ }\textbf {\bibinfo {volume} {104}},\ \bibinfo
  {pages} {054128} (\bibinfo {year} {2021})}\BibitemShut {NoStop}%
\bibitem [{\citenamefont {Bayona-Pena}\ and\ \citenamefont
  {Takahashi}(2021)}]{Pena_2021}%
  \BibitemOpen
  \bibfield  {author} {\bibinfo {author} {\bibfnamefont {P.}~\bibnamefont
  {Bayona-Pena}}\ and\ \bibinfo {author} {\bibfnamefont {K.}~\bibnamefont
  {Takahashi}},\ }\bibfield  {title} {\bibinfo {title} {Thermodynamics of a
  continuous quantum heat engine: Interplay between population and coherence},\
  }\href {https://doi.org/10.1103/PhysRevA.104.042203} {\bibfield  {journal}
  {\bibinfo  {journal} {Phys. Rev. A}\ }\textbf {\bibinfo {volume} {104}},\
  \bibinfo {pages} {042203} (\bibinfo {year} {2021})}\BibitemShut {NoStop}%
\bibitem [{\citenamefont {Boukobza}\ and\ \citenamefont
  {Tannor}(2007)}]{Boukozba2007}%
  \BibitemOpen
  \bibfield  {author} {\bibinfo {author} {\bibfnamefont {E.}~\bibnamefont
  {Boukobza}}\ and\ \bibinfo {author} {\bibfnamefont {D.~J.}\ \bibnamefont
  {Tannor}},\ }\bibfield  {title} {\bibinfo {title} {Three-level systems as
  amplifiers and attenuators: A thermodynamic analysis},\ }\href
  {https://doi.org/10.1103/PhysRevLett.98.240601} {\bibfield  {journal}
  {\bibinfo  {journal} {Phys. Rev. Lett.}\ }\textbf {\bibinfo {volume} {98}},\
  \bibinfo {pages} {240601} (\bibinfo {year} {2007})}\BibitemShut {NoStop}%
\bibitem [{\citenamefont {Linden}\ \emph {et~al.}(2010)\citenamefont {Linden},
  \citenamefont {Popescu},\ and\ \citenamefont {Skrzypczyk}}]{Linden2010}%
  \BibitemOpen
  \bibfield  {author} {\bibinfo {author} {\bibfnamefont {N.}~\bibnamefont
  {Linden}}, \bibinfo {author} {\bibfnamefont {S.}~\bibnamefont {Popescu}},\
  and\ \bibinfo {author} {\bibfnamefont {P.}~\bibnamefont {Skrzypczyk}},\
  }\bibfield  {title} {\bibinfo {title} {How small can thermal machines be?
  {T}he smallest possible refrigerator},\ }\href
  {https://doi.org/10.1103/PhysRevLett.105.130401} {\bibfield  {journal}
  {\bibinfo  {journal} {Phys. Rev. Lett.}\ }\textbf {\bibinfo {volume} {105}},\
  \bibinfo {pages} {130401} (\bibinfo {year} {2010})}\BibitemShut {NoStop}%
\bibitem [{\citenamefont {Singh}\ and\ \citenamefont
  {Johal}(2019)}]{Varinder2019}%
  \BibitemOpen
  \bibfield  {author} {\bibinfo {author} {\bibfnamefont {V.}~\bibnamefont
  {Singh}}\ and\ \bibinfo {author} {\bibfnamefont {R.~S.}\ \bibnamefont
  {Johal}},\ }\bibfield  {title} {\bibinfo {title} {Three-level laser heat
  engine at optimal performance with ecological function},\ }\href
  {https://doi.org/10.1103/PhysRevE.100.012138} {\bibfield  {journal} {\bibinfo
   {journal} {Phys. Rev. E}\ }\textbf {\bibinfo {volume} {100}},\ \bibinfo
  {pages} {012138} (\bibinfo {year} {2019})}\BibitemShut {NoStop}%
\bibitem [{\citenamefont {Singh}\ \emph {et~al.}(2020)\citenamefont {Singh},
  \citenamefont {Pandit},\ and\ \citenamefont {Johal}}]{Varinder2020}%
  \BibitemOpen
  \bibfield  {author} {\bibinfo {author} {\bibfnamefont {V.}~\bibnamefont
  {Singh}}, \bibinfo {author} {\bibfnamefont {T.}~\bibnamefont {Pandit}},\ and\
  \bibinfo {author} {\bibfnamefont {R.~S.}\ \bibnamefont {Johal}},\ }\bibfield
  {title} {\bibinfo {title} {Optimal performance of a three-level quantum
  refrigerator},\ }\href {https://doi.org/10.1103/PhysRevE.101.062121}
  {\bibfield  {journal} {\bibinfo  {journal} {Phys. Rev. E}\ }\textbf {\bibinfo
  {volume} {101}},\ \bibinfo {pages} {062121} (\bibinfo {year}
  {2020})}\BibitemShut {NoStop}%
\bibitem [{\citenamefont {Macovei}(2022)}]{Macovie_2022}%
  \BibitemOpen
  \bibfield  {author} {\bibinfo {author} {\bibfnamefont {M.~A.}\ \bibnamefont
  {Macovei}},\ }\bibfield  {title} {\bibinfo {title} {Performance of the
  collective three-level quantum thermal engine},\ }\href
  {https://doi.org/10.1103/PhysRevA.105.043708} {\bibfield  {journal} {\bibinfo
   {journal} {Phys. Rev. A}\ }\textbf {\bibinfo {volume} {105}},\ \bibinfo
  {pages} {043708} (\bibinfo {year} {2022})}\BibitemShut {NoStop}%
\bibitem [{\citenamefont {{X}iang Deng}\ \emph {et~al.}(2023)\citenamefont
  {{X}iang Deng}, \citenamefont {Shao}, \citenamefont {Liu},\ and\
  \citenamefont {Cui}}]{Deng_2023}%
  \BibitemOpen
  \bibfield  {author} {\bibinfo {author} {\bibfnamefont {G.}~\bibnamefont
  {{X}iang Deng}}, \bibinfo {author} {\bibfnamefont {W.}~\bibnamefont {Shao}},
  \bibinfo {author} {\bibfnamefont {Y.}~\bibnamefont {Liu}},\ and\ \bibinfo
  {author} {\bibfnamefont {Z.}~\bibnamefont {Cui}},\ }\bibfield  {title}
  {\bibinfo {title} {Continuous three-level quantum heat engine with high
  performance under medium temperature difference},\ }\href
  {https://doi.org/10.1063/5.0139998} {\bibfield  {journal} {\bibinfo
  {journal} {Journal of Applied Physics}\ }\textbf {\bibinfo {volume} {133}},\
  \bibinfo {pages} {124903} (\bibinfo {year} {2023})}\BibitemShut {NoStop}%
\bibitem [{\citenamefont {{Born}}\ and\ \citenamefont
  {{Fock}}(1928)}]{Adiabatic_Fock}%
  \BibitemOpen
  \bibfield  {author} {\bibinfo {author} {\bibfnamefont {M.}~\bibnamefont
  {{Born}}}\ and\ \bibinfo {author} {\bibfnamefont {V.}~\bibnamefont
  {{Fock}}},\ }\bibfield  {title} {\bibinfo {title} {{Beweis des
  Adiabatensatzes}},\ }\href {https://doi.org/10.1007/BF01343193} {\bibfield
  {journal} {\bibinfo  {journal} {Zeitschrift fur Physik}\ }\textbf {\bibinfo
  {volume} {51}},\ \bibinfo {pages} {165} (\bibinfo {year} {1928})}\BibitemShut
  {NoStop}%
%
\bibitem{mediant}
For positive real numbers $a,b,c,d >0$, the inequality $a/b > c/d$ implies
\[
 \frac{a}{b} > \frac{a+c}{b+d} > \frac{c}{d},
\]
where the fraction in the middle is known as the mediant
of the other two fractions.
\end{thebibliography}
\end{document}